\documentclass[aps,superscriptaddress,groupedaddress]{revtex4}
\usepackage{graphicx}  
\usepackage{dcolumn}   
\usepackage{bm}        
\usepackage{amssymb,amsfonts,amsmath}
\usepackage{multirow}

\def \dd {\, \text{d} }
\def \DD {\, \text{D} }
\def \zc {\, {\bar z}_{\rm c} }
\def \zcm {\, {\bar z}_{\rm c}^{\rm min}}
\def \zcmm {\, {\bar z}_{\rm c}^{\rm min}(\mu) }
\newcommand{\refeq}[1]{Eq.~(\ref{#1})}
\newcommand{\reffig}[1]{Fig.~\ref{#1}}

\begin{document}


\title{Cavity method for force transmission in jammed disordered packings of hard particles}
\date{\today}

\author{Lin Bo}\affiliation{Levich Institute and Physics Department, City College of New York, New York, NY 10031, USA}
\author{Romain Mari}\affiliation{Levich Institute and Physics Department, City College of New York, New York, NY 10031, USA}
\author{Chaoming Song}\affiliation{Physics Department, University of
  Miami, Coral Gables, FL, 33146, USA}
\author{Hern{\' a}n A. Makse}\affiliation{Levich Institute and Physics Department, City College of New York, New York, NY 10031, USA}

\begin{abstract}
The force distribution of jammed disordered packings has always been considered a central
  object in the physics of granular materials. However, many of its
  features are poorly understood. In particular, analytic relations to other key
  macroscopic properties of jammed matter, such as the contact network
  and its coordination number, are still lacking. Here we develop a
  mean-field theory for this problem, based on the consideration of the
  contact network as a random graph where the force transmission
  becomes a constraint optimization problem. We can thus use the
  cavity method developed in the last decades within the statistical
  physics of spin glasses and hard computer science problems. This method
  allows us to compute the force distribution $\text P(f)$ for random packings of
  hard particles of any shape, with or without friction. 
  We find a new signature of jamming in the small force behavior
  $\text P(f) \sim f^{\theta}$, whose exponent has attracted recent active interest: we find
  a finite value for $\text P(f=0)$, along with $\theta=0$.
  Furthermore, we relate the force distribution to a lower bound of the average coordination number $\zcmm$ of jammed packings of frictional spheres with coefficient $\mu$. This bridges the gap
  between the two known isostatic limits $\zc(\mu=0)=2D$ (in
  dimension $D$) and $\zc(\mu \to \infty)=D+1$ by extending the naive Maxwell's counting argument to frictional spheres.
  The theoretical framework describes different types of systems, such as non-spherical objects in arbitrary dimensions, 
  providing a common mean-field scenario to investigate force transmission, contact networks and coordination numbers of jammed disordered packings. \end{abstract}

\maketitle

Mechanically stable packings of granular media 
are important to a wide variety of technical
processes~\cite{behringerpowders}. 
One approach to characterize jammed granular packings is via 
the interparticle contact force network. In turn, this network 
determines the probability density of inter-particle
contact forces $\text P(f)$ and the average coordination number ${\bar
  z}$. While the force network has been studied for years, there is yet no unified
theoretical framework to explain the common observations in granular packings,
ranging from frictional to frictionless systems, from spherical to non-spherical particles and
in any dimensions. 

Experimental force measurements~\cite{Zhou16062006, Liu28071995,
  PhysRevE.57.3164, PhysRevE.60.5872,Erikson2002, Brujic2003,
  Brujic2003b} and
simulations~\cite{Radjai1996,OHern2001,PhysRevE.62.2510,Makse2000}
have shown that the interparticle forces are inhomogeneously
distributed with common features: $\text P(f)$ near the jamming
transition has a peak at small forces and an approximate exponential
tail in the limit of large $f$.  It is argued that the development of
a peak is a signature of the jamming transition~\cite{OHern2001}.  The
scaling of $\text P(f)$ in the limit of small forces is directly
related to the mechanical stability of the
packing~\cite{PhysRevLett.109.125502, lerner2013low}: the contacts
bearing small loads are the ones for which a local buckling is the
most easily achieved~\cite{lerner2013low}, and they must thus
exist in small enough density for the packing to be stable.  Some parts
of the qualitative behavior of $\text P(f)$ are correctly captured by
simplified mean-field models, such as the
q-model~\cite{Liu28071995,PhysRevE.53.4673} and Edwards'
model~\cite{Brujic2003, Brujic2003b}. Both of them describe the
exponential decay at large force and a power law behavior $\text
P(f)\sim f^{\theta}$ for small forces~\cite{Brujic2003, Brujic2003b,
  socolar1998average}. However, the exponent $\theta$ obtained by the
q-model (an integer $\theta \geq 1$~\cite{Liu28071995,
  socolar1998average}) is larger than the one obtained by recent numerical
simulations accessing the low force limit with increasing accuracy $0.2
\lesssim \theta \lesssim
0.5$~\cite{PhysRevLett.109.205501,lerner2013low},
while the exponent obtained by Edwards' model $\theta = 1/({\bar
  z}-2)$ depends strongly on the dimension (through ${\bar z}$), when
simulations show it does not~\cite{PhysRevLett.109.205501}. A recent
replica calculation finds an exponent $\theta \approx 0.42$ in
infinite dimensions of
space~\cite{kurchan_exact_2013,charbonneau_exact_2013,charbonneau_fractal_2014}.
Other theoretical approaches based on entropy maximization similar to
Edwards' statistical mechanics~\cite{Edwards1989} also recover the
large force exponential
decay~\cite{Kruyt2002,bagi2003statistical,Goddard20045851,PhysRevE.79.061301}.
Some of those works which simulated the force network
ensemble~\cite{PhysRevLett.92.054302} on strongly hyperstatic ${\bar
  z}=6 > \zc$ triangular lattice~\cite{PhysRevLett.100.238001} and on
2-D (3-D) disordered hyperstatic frictionless bead packings generated
using molecular dynamics~\cite{van2007tail} however predict a force
distribution that decays faster than exponential, and so does a
mean-field theory based on replica theory of spin
glasses~\cite{PhysRevLett.109.205501,RevModPhys.82.789,berthier2011microscopic}.
Recent measurements in the bulk of
packings~\cite{Zhou16062006,Brujic2003, Brujic2003b, Majmudar2005}
find downward curvature on semi-log plot of $\text P(f)$.

The average coordination number per particle ${\bar z}$ is
another key signature of jamming. Close to jamming, 
many observables (pressure, volume fraction, shear modulus
or viscosity~\cite{lerner2012toward} to name
a few) scale with the distance ${\bar z}-\zc$ to the average coordination number $\zc$ at the
transition. Understanding the value of $\zc$ is
thus of primary importance. The case of frictionless spheres, for which, $\zc= 2D$ where $D$
 is the dimension, has been rationalized based on counting
arguments leading to the isostatic conjecture~\cite{Alexander199865,PhysRevE.60.687,PhysRevLett.81.1634}:
a lower bound $\zc \geq 2D$ is
provided by Maxwell's stability
argument~\cite{maxwell1864calculation}, and an upper bound $\zc \leq
2D$ is given by the geometric constraint of having the
particles exactly at contact, without overlap. The problem, however,
turns out to be more complicated when friction is considered, and
no method is known so far to predict $\zc$ in such a case. The (naive)
generalization of the counting arguments gives the bounds $D+1 \leq \zc
\leq 2D$ in frictional packings, independently of the value of the interparticle
friction coefficient $\mu$. Indeed, there is a range of $\zc$ obtained numerically and
experimentally~\cite{PhysRevLett.92.054302,PhysRevLett.81.1634,PhysRevE.65.031304,Silbert2002,PhysRevE.72.011301,PhysRevE.75.010301,song2008phase,Wang20103972,C001973A,
ohern2013,brujic2007measuring}. However,
for small $\mu$, this range never extends to the predicted lower
bound, as packings with low friction coefficient lie close to $\zc = 2D$.

Here we present a theoretical framework at a mean-field level to 
consider force transmission as a constraint optimization problem on random graphs, and
study this problem with standard tools, namely the
cavity method~\cite{M.Mezard1986}. We first obtain the force distribution
for spheres, in frictionless and frictional cases, both in
two and three dimensions.  Besides showing the experimentally known 
approximate exponential fall-off at large forces, these
distributions bring a new insight in the much less known small force
regime. In particular, we find for frictionless spheres in both 2 and
3 dimensions a finite value for $\mathrm{P}(f)$ in $f=0$, leading to
a mean-field exponent $\theta = 0$ for small forces.

Our framework is also the first one to show theoretically
how the frictional coefficient $\mu$ affects the average number of
contacting neighbors $\zc$ at the jamming transition, and we find a
lower bound $\zcmm$ for this number (which is also a lower bound on
${\bar z}$, since ${\bar z} \geq \zc$ in a packing). We achieve this
by generalizing in a careful way the naive Maxwell counting arguments,
considering the satisfiability of force and torque balances
equations. Linking $\zc$ to the behavior of force and torque balances
is not a new idea, as it was already suggested by Silbert {\it et
  al}.~\cite{Silbert2002}. Furthermore the generalized isostaticity
picture~\cite{PhysRevE.75.010301} gives a bound on the number of fully
mobilized forces (\textit{ie} the number of tangential forces which
are taking the maximally value allowed by the Coulomb law) based on
the value of $\zc$. However, none of these works derive a bound for
$\zc$ itself, at a given $\mu$. The bound $\zcmm$ that we obtain
interpolates smoothly between the two isostatic limits at $\mu =0$ and
$\mu \to \infty$.

\section{Force and torque balances as a satisfiability problem}

A packing can be described as an ensemble of particles with given
position and orientation, having interparticle contacts such that
(\textit{i}) there is no overlap between particles, (\textit{ii}) force 
and torque balances are satisfied on
every particle, and
(\textit{iii}) the packing is rigid, \textit{ie} there is no
displacement of particles, individual or collective, that does not
create a violation of conditions (\textit{i}) or (\textit{ii})
(except for global translation or rotation of the packing). In this work, we
consider the geometric configuration as given (\emph{ie} condition (\textit{i})
is fulfilled and taken for granted), hence we do not discuss about
the spatial degrees of freedom but rather focus on the force degrees
of freedom.

Now, let us argue that, in the case of random packings of spheres
under finite pressure,
fulfilling condition (\textit{ii}) (force and torque balances) implies
that condition (\textit{iii}) (stability) is also fulfilled.
Force and torque balances under finite pressure ensure (by definition)
that the system is stable against isotropic
compression. The average force in the packing is then proportional to
the pressure. But, as such, nothing is guaranteed concerning the stability of the
packing against more general perturbations, as defined by the
condition (\textit{iii}). 
In the case of
sphere packings, however, it is very likely that packings verifying
 force and torque balances while still being mechanically unstable have
some sort of order, at least locally (one can think of the case of a
row of frictionless spheres perfectly aligned for example, which can
maintain force balance, but is unstable). This is what is observed at
the jamming transition, when packings prepared under isotropic
pressure are naturally stable~\cite{OHern2003}. Hence, for
random packings of spheres, the requirement of force and torque balances is argued
to be enough for having stability too, \textit{ie} conditions
(\textit{ii}) and (\textit{iii}) are simultaneously fulfilled.

Similarly to the global rigidity condition, force and torque balances are
entirely constrained by the contact network of the packing. If we
define $\vec{r}^{\,a}_{i}$ as the vector joining the center of
particle $a$ and the contact $i$ on it, the contact network is
uniquely defined by the complete set 
$\{\vec{r}^{\,a}_{i}\}$. Calling $\vec{f}^{\,a}_i$ the force acting
on particle $a$ from contact $i$, force and torque balances read:
\begin{equation}
  \label{eq:force_balance}
  \begin{array}{llr}
    \displaystyle \sum_{i \in \partial a} \vec{f}^{\,a}_i  = 0, 
     \qquad &  \vec{r}^{\,a}_i \cdot \vec{f}^{\,a}_i < 0, & \multirow{2}{*}[-0.8em]{$\qquad  \forall \, a$}\\  
    \displaystyle \sum_{i \in \partial a} \vec{r}^{\,a}_i \times \vec{f}^{\,a}_i  =  0, \qquad &
       f_{i}^{\,t} \leq \mu f_{i}^{\,n}, &
  \end{array}
\end{equation}
where the notation $\partial a$ denotes the set of contacts
of particle $a$. We explicitly take into account friction by
decomposing the force into normal and tangential parts $\vec{f}^{\,a}_i= -
f_{i}^{\,n} \hat{n}^{\,a}_i + f_{i}^{\,t} \hat{t}^{\,a}_i$, 
where $\hat{n}^{\,a}_i$ and $\hat{t}^{\,a}_i$ are normal and tangential unit vectors
 to the contact, respectively. The inequality $\vec{r}^{\,a}_i \cdot \vec{f}^{\,a}_i < 0$ ensures the repulsive nature of
the normal force. The last inequality ensures Coulomb's law with
friction coefficient $\mu$. 

\begin{figure}[!t]
\includegraphics[width=0.618\textwidth]{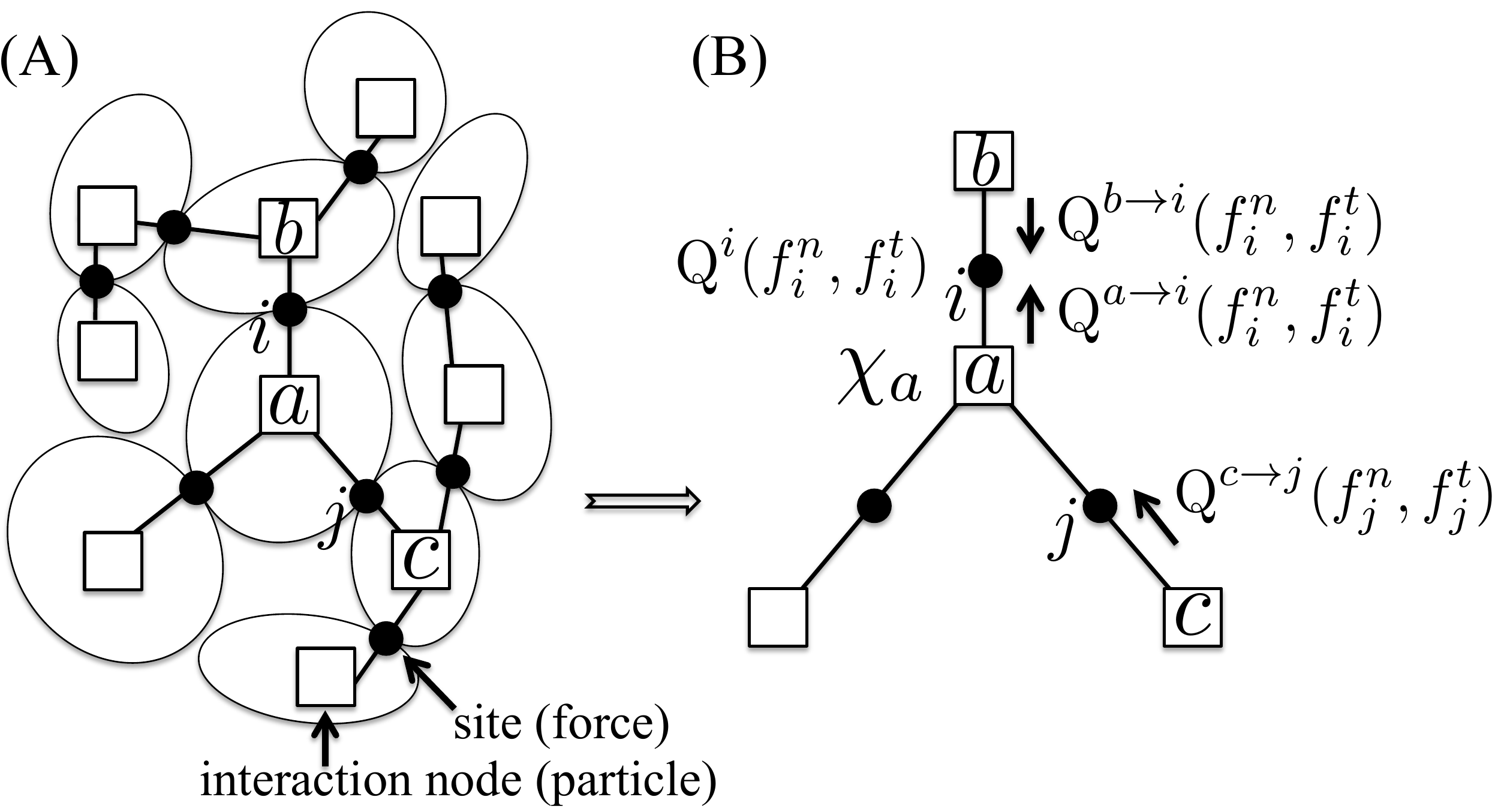}
\caption{{\bf Factor graph and variables}. (A) Building the factor graph of contacts from a packing~\cite{Mezard2009}. The contacts
  becomes sites (filled circles) with associated variables being
  forces. The particles are reduced to interaction nodes (open squares) dealing with
  force/torque balance~\refeq{eq:force_balance}.
  (B) Part of factor graph used to compute $\text{Q}^i$ and $\text{Q}^{a\to i}$ by using \refeq{eq:P2Pto} and \refeq{eq:cavity_eq}. On the edges around particle $a$, the arrows indicate that we consider the force probabilities $\text{Q}^{\to}$ as `messages' going through contacts in the contact network.  In our algorithm, $\text{Q}^{a\to i}$ is iteratively updated by the uncorrelated force probability $\text{Q}^{c\to j}$ on neighboring edges with the force/torque balance constraint $\chi_a(\{ f^{\,n},f^{\,t},\hat{n}^{\,a},\hat{t}^{\,a} \}_{\partial a})$ on that particle $a$, verifying the cavity equation \refeq{eq:cavity_eq}. The marginal probability $\text{Q}^i(f_i^{\,n},f_i^{\,t})$ on site $i$ is calculated as a product of $\text{Q}^{a\to i}$ and $\text{Q}^{b\to i}$ as in \refeq{eq:P2Pto}.  }\label{fig:from_network_to_graph} 
\end{figure}

The constraints induced by force and torque balances on the forces
$\vec{f}_{i}$ are not always satisfiable. In the case of frictionless
spheres, we can recover the known ${\bar z} \geq 2D$ directly from
force balance alone. The naive Maxwell counting
argument~\cite{PhysRevLett.81.1634} applied to frictionless spheres
reduces~\refeq{eq:force_balance} to a set of linear equations by taking into account only force balance and neglecting the repulsive
nature of the forces. Maxwell argument considers the minimal number of
forces needed to satisfy \refeq{eq:force_balance} which gives, per sphere,
$D$ equations and ${\bar z}/2$ variables (forces), implying ${\bar z} \geq 2D$ to have a solution. Below this threshold, there is
generically no solution to \refeq{eq:force_balance}. 
To accurately extend this
counting argument to more general conditions (frictional, repulsive, and/or
non-spherical particles), one must take into account all the
constraints in \refeq{eq:force_balance}, including the repulsive
nature of the forces and the Coulomb condition for frictional
packings. Indeed, the naive Maxwell argument, neglecting those constraints,
concludes ${\bar z}> \, \zcmm=D+1$ for any frictional
packing of spheres, ignoring the dependance on the friction coefficient~\cite{edwards1999statistical}.
On the other hand, below we show that including the above mentioned constraints
we obtain
an accurate lower bound $\zcmm$, explicitely depending on $\mu$.

We tackle the problem of satisfiability of force and torque balances \refeq{eq:force_balance} by
looking at the contact network in an amorphous packing as an instance
of \emph{random graph}. As depicted in
\reffig{fig:from_network_to_graph}A, starting from a packing of
$N$ particles, we explicitly construct a so-called factor graph~\cite{Mezard2009},
considering the $M={\bar z}N/2$ contacts as `sites', and the $N$
particles as `interaction nodes'. Each site $i$ bears two vectors
$\vec{r}^{\,a}_i$ and $\vec{r}^{\,b}_i$ and two opposite forces
$\vec{f}^{\,a}_i = -\vec{f}^{\,b}_i$ (one per particle involved in the
contact). Note that $\{\hat{n}^{\,a}_i\}$ are
uniquely determined by the contact network $\{\vec{r}^{\,a}_i\}$ and represents the `quenched' disorder in the system,  whereas $\{\hat{t}^{\,a}_i\}$ are free to rotate in the plane tangent to contact.
On each interaction node (particle) $a$, we enforce
force balance, torque balance, repulsive interactions and Coulomb friction conditions on its
$z_a$ neighboring sites by an interaction function,
\begin{equation}\label{eq:interaction_node}
  \begin{array}{rl}
  \chi_a(\{ f^{\,n},f^{\,t},\hat{n}^{\,a},\hat{t}^{\,a} \}_{\partial a}) = & \! \! \! \! \delta
  \Big( \displaystyle \sum_{i \in \partial a}  \vec{f}^{\,a}_i \Big)\: \delta
  \Big( \displaystyle  \sum_{i \in \partial a}  \vec{r}^{\,a}_i \times \vec{f}^{\,a}_i  \Big)\\
  & \quad \times \displaystyle \prod_{i \in \partial a} \Theta(f_{i}^{\,n}) \Theta(\mu
  f_{i}^{\,n} - f_{i}^{\,t}).
  \end{array}
\end{equation}
We define the partition function $Z$ and entropy $S$ for the
problem of satisfiability of force and torque balances for a fixed realization of 
the quenched disorder $\{\hat{n}^{\,a}_i\}$ (as shown in Appendix):
\begin{equation}
  \begin{array}{rl}
  e^S =  Z = & \displaystyle \int \prod_{i=1}^M \dd f_{i}^{\,n} \dd f_{i}^{\,t} 
   \dd  \hat{t}^{\,a}_i \dd \hat{t}^{\,b}_i  \, \delta(\hat{t}^{\,a}_i
   + \hat{t}^{\,b}_i)\\ 
   & \hskip-0.2in \displaystyle \times \,\delta(\sum_{i=1}^M f_{i}^{\,n} - Mp)
 \prod_{a=1}^N \chi_a(\{ f^{\,n},f^{\,t},\hat{n}^{\,a},\hat{t}^{\,a} \}_{\partial a}).
  \end{array}
  \label{eq:partition_function}
\end{equation}
Without loss of generality, we work in the constant pressure $p$
ensemble: if we find a solution to the force and torque balances
problem having a pressure $p'$, we can always find one solution with pressure
$p$ by multiplying all forces by $p/p'$. Note that the only effect of
the applied pressure in a hard sphere packing at zero temperature is
to set the average normal force. The contact network is unmodified by
a change in pressure as the particles are hard.

If the entropy is finite, there exists a solution to force and torque balances. The
satisfiability/unsatisfiability threshold of force and torque balances
is the coordination number $\zcmm$ that separates the region of finite $S$ from the region for which $S\to -\infty$, corresponding to an underdetermined/overdetermined set of \refeq{eq:force_balance}, respectively. 

Within this framework, the satisfiability problem \refeq{eq:force_balance}
 is one of the well studied class of constraint satisfiability
problems (CSP) defined on random networks~\cite{Mezard2009}. These problems are ubiquitous
 in statistical physics and computer science and have
attracted a lot of attention in recent years. Thus powerful
methods from statistical physics have been developed to study
them~\cite{Mezard2009}.

We will work here with random graphs, retaining from actual packings
the distribution of coordination number $\text{R}(z)$.  In this work,
we use as $\text{R}(z)$ a truncated Gaussian distribution, more
precisely, $\text{R}(2<z<6) \propto e^{-(z-z_0)^2/2}$ in 2-D and
$\text{R}(3<z<9) \propto e^{-(z-z_0)^2/2.88}$ in 3-D, both providing
very good fit to experimental and numerical
data~\cite{henkes_local_2010,PhysRevE.65.031304,WangP2010,PhysRevE.72.011301}. (Note
that due to the truncation $z_0\neq \bar{z}$, we choose $z_0$ to
achieve a desired $\bar{z}$.) We clarify that although this choice of
$\text{R}(z)$ relies on quantities extracted from previous numerical
studies, most of the results we obtain on the force distribution are
insensitive to the exact form of $\text{R}(z)$. The main change observed for
frictionless sphere packings is the behavior near the peak of the
force distribution, as is discussed in Sec.~\ref{sec:pofF}.

For the distribution of contacts around one
particle $\Omega(\vec{r}^{\,a}_1,\dots, \vec{r}^{\,a}_{z_a})$, we use
a flat measure over all contacts positions that do not create overlaps
between the $z_a$ corresponding particles. For spherical particles this
distribution is $\Omega(\vec{r}^{\,a}_1,\dots,
\vec{r}^{\,a}_{z_a})=\prod_{i\neq j\in [1,z_a]} \Theta(1/2 -
\vec{r}^{\,a}_i.\vec{r}^{\,a}_j)$ with $\Theta$ the Heaviside function. 
A real contact network of a two or three-dimensional packing shows
some finite dimensional structure, of course, 
and treating it as random graphs can only be
an approximation. This amounts to a mean-field approximation,
neglecting correlations between the different contacting forces acting
on one particle. This approximation is routinely used in the context of
spin-glasses for example~\cite{M.Mezard1986}. From this point of view, we stand on the
same ground as the q-model
approach~\cite{Liu28071995,PhysRevE.53.4673}, and the Edwards'
approach of Bruji{\'c} {\it et al}.~\cite{Brujic2003, Brujic2003b,makse2004statistical}.

Even though we keep a part of the finite dimensional geometric
constraints through the distributions $\Omega(\vec{r}^{\,a}_1,\dots,
\vec{r}^{\,a}_{z_a})$ and $\text{R}(z)$, the mean-field approach
neglecting correlations between neighboring contact forces should only
be exact in the limit of high dimension, where we expect those
correlations to vanish (some recent progress however suggest that
correlations are not completely trivialized by the high dimensionality
in packings, due to a fullRSB phase transition close to
jamming~\cite{charbonneau_fractal_2014}). In finite dimension, it is
clear that short range correlations matter, and having a mean-field
approximation can only be a first step. One finite dimensional feature
that we miss is the detailed link between local structure and force
distribution. A stability analysis set constraints relating the
detailed features of the near contact structure and the force
distribution at low forces~\cite{PhysRevLett.109.125502,
  lerner2013low}. The mean-field computation we are providing here
does not include that constraint, and there may be some finite
dimensional effects at low forces that modify the mean-field picture
we are giving here, as shown in the result section~\ref{sec:pofF}.

\section{Cavity Method}
\subsection{General formalism}

As the simplest case, we restrict the description of this section to
packings of spheres, with obvious generalization to non-spherical
objects. For a given disordered packing, each particle $a$ has unique
surroundings, different from its neighbors or other particles in the
packing. These surroundings are defined by the contact number $z_a$
and contact vectors $\{\vec{r}_i^{\, a}\}$.  If the system is
underdetermined, several sets of forces in the system satisfy force
and torque balances, and each contact force has a certain probability
distribution $\text{Q}^i(f_i^{\,n},f_i^{\,t})$.

The local disorder makes each contact
unique, and the probability distributions of forces $\text{Q}^i(f_i^{\,n},f_i^{\,t})$ are different from
contact to contact.  We define the overall force distribution in a
packing $\text{P}(f^{\,n},f^{\,t})$ as an average over the probability
distributions of forces over the contacts:
\begin{equation}
  \label{eq:average_P}
  \text{P}(f^{\,n},f^{\,t}) \equiv \langle \text{Q}^i(f^{\,n},f^{\,t}) \rangle = \frac{1}{M} \sum_i \text{Q}^i(f^{\,n},f^{\,t}).
\end{equation}
\noindent Next, we show that on a random graph, we can access the
distributions $\text{Q}^i(f_i^{\,n},f_i^{\,t})$ with a self-consistent set of
local equations using the cavity method~\cite{Mezard2009}. In this
description, we work at fixed pressure $p$, \textit{ie} we consider any two solutions differing
only by an overall rescaling of the pressure to be only one genuine
solution.
Each contact is
linked to two particles, $a$ and $b$.
We denote
$\text{Q}^{a\to i}(f_i^{\,n},f_i^{\,t})$ the probability distribution of
the force $\vec{f}_i$ of a site (contact) $i$, if $i$ is
connected only to the interaction node (particle) $a$, that is, if we
remove particle $b$ (dig a cavity) from the packing. The main assumption of
the cavity method is to consider that $\text{Q}^{a\to i}(f_i^{\,n},f_i^{\,t})$ and $\text{Q}^{b\to
  i}(f_i^{\,n},f_i^{\,t})$ are uncorrelated. Therefore, we can write the
probability of forces at contact $i$ as:
\begin{equation}
  \label{eq:P2Pto}
  \text{Q}^i(f_i^{\,n},f_i^{\,t}) = \frac{1}{Z^i} \text{Q}^{a\to
    i}(f_i^{\,n},f_i^{\,t})\, \text{Q}^{b\to i}(f_i^{\,n},f_i^{\,t}), \,\, \{a,b\}=\partial i
\end{equation}
with $Z^i$ the normalization.

Under the mean-field assumption a set of local equations (called cavity equations)
relates the $\text{Q}^{\to}$'s, as depicted in \reffig{fig:from_network_to_graph}B:
\begin{equation}
  \label{eq:cavity_eq}
  \begin{array}{rl}
    \text{Q}^{a\to i}(f_i^{\,n},f_i^{\,t})  = & \displaystyle \frac{1}{Z^{a \to i}} \displaystyle \int \dd  \hat{t}_i^{\,a}
    \displaystyle \prod_{j\in \partial a - i} \dd f_j^{\,n} \dd f_j^{\,t}\dd \hat{t}_j^{\,a}\,
     \,  \\& \hskip-0.35in   \displaystyle \times \chi_a(\{ f^{\,n},f^{\,t},\hat{n}^{\,a},\hat{t}^{\,a} \}_{ \partial
       a}) \! \! \prod_{c  = \partial j - a }\! \! \! \! \text{Q}^{c\to j}(f_j^{\,n},f_j^{\,t})
     \\
     \equiv & \mathcal{F}_{a\to i} \big( \big\{ \text{Q}^{c \to j}\big\} \big),
    \end{array}
\end{equation}
where the notation $\partial x - y$ stands for the set of neighbors of
$x$ on the graph except $y$, and $Z^{a \to i}$ is the
normalization. Crucially, we do
not average over the contact directions $\{\hat{n}^{\,a}\}_{\partial a}$ at this stage
(whereas the q-model~\cite{Liu28071995,PhysRevE.53.4673} and Edwards' model~\cite{Brujic2003, Brujic2003b} do). This implies that every link $a \to i$ has
a different distribution, due to the local `quenched' disorder
provided by the
 contact network $\{\hat{n}^{\,a}\}_{\partial a}$ and contact number $z_a$.
Hence, finding a set of $\text{Q}^{\to}$ that are solutions of
\refeq{eq:cavity_eq} allows to get the distribution of forces on every
contact \textit{individually}, through the use of \refeq{eq:P2Pto}.

Looking for a solution of \refeq{eq:cavity_eq} for a given instance of
the contact directions (meaning for one given packing) is
possible. These equations are commonly encountered as `cavity
equations' in the context of spin glasses or optimization problems
defined on random graphs~\cite{Mezard2009}, and they can be solved by
message passing algorithms like Belief Propagation. Here we follow another route, 
since we are interested in $\text{P}(f^{\,n},f^{\,t})$ for not only one
packing but over the ensemble of all
random packings. Thus, we study the solutions of the cavity equations in the thermodynamic limit to provide typical solutions for large packings. As
in statistical mechanics, the partition function will be
dominated by the relevant typical configurations which we
expect will be realized in experiments.

The set of cavity equations \refeq{eq:cavity_eq} might \textit{a priori} admit several solutions. 
However, at the satisfiability/unsatisfiability threshold, there is
only one solution to force balance \refeq{eq:force_balance}, which
means only one solution to \refeq{eq:cavity_eq} (a $\delta$-function
on every site, centered on the solution of force and torque balances).  To get
this threshold, we may thus take for granted that there is a unique
solution to the cavity equations, a case known as replica symmetric
(RS) in the spin glass terminology. This assumption will be fully
justified \textit{a posteriori} by the fact that we recover the
correct threshold in the known case of frictionless spheres.

  In the thermodynamic limit, the set of $\text{Q}^{\to}$'s that are
  solutions of \refeq{eq:cavity_eq} are distributed according to the
  probability $\mathcal{Q}(\text{Q}^{\to})$:
\begin{equation}
  \label{eq:def_QofQ}
  \mathcal{Q}(\text{Q}^{\to}) \equiv \frac{1}{2M} \sum_{a\to i} \delta \left[
    \text{Q}^{\to} - \text{Q}^{a\to i} \right] 
\end{equation}
In this case, we can replace the sum over $a\!\! \to\!\! i$ by a
continuum description of the $\text{Q}^{\to}$'s based on their
distribution $\mathcal{Q}(\text{Q}^{\to})$. The probability that a
given $\text{Q}^{\to}$ is set by a cavity equation
\refeq{eq:cavity_eq} involving $z-1$ contact is proportional to $z\, \text{R}(z)\,
\Omega(\hat{n}_i,\{\hat{n}_j\})$. Thus, averaging over the ensemble of random
graphs, \refeq{eq:def_QofQ} becomes a self-consistent 
equation~\cite{Mezard2009,M.Mezard1986}:
\begin{equation}\label{eq:cavity_averaged}
  \begin{array}{rl}
  \mathcal{Q} (\text{Q}^{\to}) = & \! \! \! \! \displaystyle  \frac{1}{\mathcal{Z}}\sum_{z} z \, \text{R}(z)
    \int \Omega(\hat{n},\{\hat{n}_j\}) \prod_{j=1}^{z-1} \dd \hat{n}_j   \\
    & \qquad \times \displaystyle  \DD \text{Q}^{\to j}  
    \: \mathcal{Q}(\text{Q}^{\to j} )
    \delta \Big[\text{Q}^{\to}- \mathcal{F}_{\to} \big( \big\{
    \text{Q}^{ \to j} \big\} \big) \Big].
  \end{array}
\end{equation}
where $\mathcal{Z}$ is the normalization. Note that the value of the
integral does not depend on the choice of $\hat{n}$. 
\noindent Once a solution to \refeq{eq:cavity_averaged} is known, we  deduce
the force distribution $\text{P}(f^{\,n},f^{\,t})$ in the overall packing as
the average of all these probability distributions and contacts:
\begin{equation}
  \begin{array}{rl}
  \text{P}(f^{\,n},f^{\,t}) & = \langle \text{Q}^i(f^{\,n},f^{\,t}) \rangle \\
  & = \displaystyle \frac{1}{\mathrm{Z}_{\mathrm{P}}} \left [\int 
  \DD \text{Q}^{\to }  \mathcal{Q}(\text{Q}^{\to }) 
\text{Q}^{\to }(f^{\,n},f^{\,t}) \right]^2
  \end{array}
\label{fdis}
\end{equation}
where $\mathrm{Z}_{\mathrm{P}}$ is the normalization to ensure $\int \text{P}(f^{\,n},f^{\,t}) = 1$.

Equation~(\ref{eq:cavity_averaged}) stands out as the main and crucial
difference with previous approaches, in particular the q-model~\cite{Liu28071995,PhysRevE.53.4673} and
Edwards' description~\cite{Brujic2003, Brujic2003b}. 
Although these approaches also neglect correlations, our work does not
reduce to those models
 due to a fundamentally different way of treating the disorder
in the packing. Here, we consider a site-dependent
$\text{Q}^i(f_i^{\,n},f_i^{\,t})$, where the Edwards' model and q-model create all sites
equal. Thus in our method the average over the packing configurations is
not done at the same level as the average over forces. That is, 
we perform a quenched average over the disorder of the graph.  
As random packings in two or three dimensions have a rather small
connectivity, the fluctuations in the environment of one particle are
large: no particle stands in a `typical' surrounding. Hence, the
average over the `quenched' disorder (the packing configurations) must be done
with care. Averaging directly~\refeq{eq:cavity_eq} (a so-called
`annealed' average in spin-glass terminology), as the previously cited
approaches do~\cite{Liu28071995,Brujic2003, Brujic2003b,PhysRevE.53.4673}, amounts to
neglect the site to site fluctuations. Performing a `quenched'
average as in~\refeq{eq:cavity_averaged}, however, allows to take into account these fluctuations
correctly~\cite{M.Mezard1986}, and leads to a force distribution which
is the average force distribution over the ensemble of possible
packings, as opposed to the force distribution of an averaged
packing. 

This issue becomes also crucial to the study
of the satisfiability transition of force and torque balances. For example, the
q-model describes a force distribution in a packing of frictionless
spheres (\textit{ie} it finds solution to force balance), even in
cases where we know there is no solution, such as
when ${\bar z}<6$ in 3-D frictionless systems. This can be understood by looking at the entropy
defined in \refeq{eq:partition_function}. The annealed average over
disorder done in the q-model amounts to compute the averaged partition function
$\overline{Z}$, and get the entropy through $S_{\text{an}}=\ln
\overline{Z}$, with $Z$ defined in \refeq{eq:partition_function}. 
But one can show that $\overline{Z}$ is always
finite for ${\bar z}\geq 2$. Indeed, for ${\bar z}=2$, an infinite
row of perfectly aligned spheres will satisfy force and torque balances
and will contribute to the partition function. A straightforward
generalization of this example shows that for ${\bar z}\geq 2$, the
annealed entropy is finite. On the contrary, the quenched average
amounts to compute the averaged entropy $S_{\text{qu}} = \overline{\ln
  Z}$. Now, for our frictionless sphere example, typical
configurations with ${\bar z}<6$ cannot satisfy force balance, and
their diverging negative entropy will dominate the average $S_{\text{qu}}$. Hence, the quenched
average correctly captures the satisfiability/unsatisfiability transition at $\bar{z}=6$ while the
annealed average does not.

Equation~(\ref{eq:cavity_averaged}) is typically hard to solve,
since it is a self-consistent equation for a distribution of distributions $\mathcal{Q}(\text{Q}^{\to})$. 
For this purpose, we use  a Population Dynamics algorithm (shown in the next section),
familiar to optimization problems \cite{Mezard2009}. This
method consists to describe the distribution $\mathcal{Q}$ via
a discrete sampling (a `population') made of a large number of
distributions $\text{Q}^{\to}$. Applying iteratively
\refeq{eq:cavity_averaged}, we find, if it exists, a fixed-point
of the distribution $\mathcal{Q}(\text{Q}^{\to})$. 

It is interesting to discuss the different types of solutions expected
from \refeq{eq:cavity_averaged}.
 For a given contact network, if the system is satisfiable and underdetermined, hence admits an
infinite set of solutions for force and torque balances, the distributions
$\text{Q}^i(f^{\,n},f^{\,t})$ should be
broad, allowing each force to take values in a non-vanishing
range. This means that on each contact, $\text{Q}^{a\to i}(f^{\,n},f^{\,t})$
and $\text{Q}^{b\to i}(f^{\,n},f^{\,t})$ should be broad and overlapping. If
the system is neither under- nor overdetermined (\textit{ie} isostatic),
there is only one solution to force and torque balances for every site $i$,
and each $\text{Q}^i(f^{\,n},f^{\,t})$ is a Dirac $\delta$-function
centered on the solution. If the system is overdetermined or unsatisfiable, there is
typically no solution to \refeq{eq:cavity_averaged}, meaning that an
algorithm designed to solve it would not converge. In practice, since
we perform a population dynamics algorithm to average over all possible
packings, if one starts with a set $\{\text{Q}^{a\to i}(f^{\,n},f^{\,t})\}$ of broad distributions as a
guess for the solution, both isostatic and overdetermined ensembles
show that all $\{\text{Q}^{a\to i}(f^{\,n},f^{\,t})\}$ shrink to $\delta$-distributions after a few iterations,
while underdetermined ensemble always gives broad (not vanishing)
probabilities. Therefore the threshold $\zcmm$ of the
satisfiability/unsatisfiability transition for force transmission can
be located by measuring the width of the force distributions
$\{\text{Q}^{a\to i}(f^{\,n},f^{\,t})\}$.
  
 The location of this transition, in turn, constitutes a lower
bound for the possible coordination number $\zcmm$, which extends Maxwell's counting
argument for $\mu =0$ to any friction. An additional quantity available is the force distribution
itself, as a function of $\zcmm$. Therefore, our approach
explicitly relates two essential properties of the jamming
transition: the average coordination number and the force distribution.

\subsection{Population Dynamics algorithm}
\label{sec:PD}

As discussed in the previous section, satisfiability of force/torque balance
is studied via the behavior of the probability distribution
$\mathcal{Q}(\text{Q}^{\to})$ of the distributions
$\text{Q}^{\to}(f^{\,n}, f^{\,t})$,
defined by \refeq{eq:def_QofQ}. This quantity is obtained for a
given graph ensemble, defined by the distribution of connectivity
(contact number) $\mathrm{R}(z)$ and the joint probability
distribution of the contact directions $\Omega(\{\hat{n}\})$ around
every interaction node (particle), via the cavity
equations~\refeq{eq:cavity_eq} and ~\refeq{eq:cavity_averaged}.

Equation (8) is a self-consistent equation for
$\mathcal{Q}(\text{Q}^{\to})$. Self-consistent equations can generally
be solved by an iterative algorithm, and this is the method we use
here. However, there is one difficulty arising from the fact that
$\mathcal{Q}(\text{Q}^{\to})$, which is a distribution of
distributions, is a complex object in itself and is hard to simply
describe in a computer program. This is a recurring problem of
solving cavity equations, and the solution developed to overcome this
difficulty is called Population Dynamics algorithm. 

Population Dynamics algorithm solves the problem of representing
$\mathcal{Q}(\text{Q}^{\to})$ by describing it by a large sample drawn
from $\mathcal{Q}$. This sample $\{\text{Q}^{\to}_1, ...,
\text{Q}^{\to}_m\}$ is called a `population', and it matches
$\mathcal{Q}(\text{Q}^{\to})$ in the sense that it contains more
elements close (in a function norm sense) to $\text{X}_1$ than
$\text{X}_2$ if $\mathcal{Q}(\text{X}_1) >
\mathcal{Q}(\text{X}_2)$. More precisely, the approximation of
$\mathcal{Q}(\text{Q}^{\to})$ it gives is $\frac{1}{m}\sum_{i} \delta
\left[ \text{Q}^{\to} - \text{Q}^{\to}_i \right]$. This population of
course needs to be as large as possible in order to give an accurate
description of $\mathcal{Q}(\text{Q}^{\to})$. Now, with this change of
description, the algorithm needs to specify an iterative method (a
`dynamics') to
make the population converge to the solution of~\refeq{eq:cavity_averaged}. This is
the heart of the  Population Dynamics algorithm.

One starts from an initial population guess  $\{\text{Q}^{\to, (0)}_1, ..., \text{Q}^{\to, (0)}_m \}$ 
at iterative time $t=0$. The only requirements on those initial distributions are (a) repulsive normal
force condition $\text{Q}^{\to, (0)}_i(f^{\,n}<0, f^{\,t}) = 0$, (b) Coulomb
friction condition $\text{Q}^{\to, (0)}_i(f^{\,n}, f^{\,t}>\mu f^{\,n}) = 0$ and
(c) fixed pressure $ mp = \sum_i \int \dd f^{\,n} \dd f^{\,t} \, f^{\,n}\, \text{Q}^{\to,
  (0)}_i(f^{\,n}, f^{\,t})$. In practice, we tested several initial population
guesses (Gaussian, Gaussian plus random noise, uniform on the region $[0,
(3p/\mu)^{1/3}]\times [0, \mu f^{\,n}]$ of the $f^{\,n}, f^{\,t}$ plane), with
identical results at the end of the iteration procedure. One then
iterates in the following way at time $t$, starting from $t=0$:

(1) Pick randomly a distribution $\text{Q}^{\to, (t)}_i$ in the current population.

(2) Pick a contact number $z$ within the distribution $\frac{z \:
  \text{R}(z)}{\sum_{z} z \: \text{R}(z)}$, and generate a set of $z$
random contact directions $\hat{n},\hat{n}_1,...,\hat{n}_{z-1}$ with the distribution $\Omega(\{\hat{n}\})$.

(3) Pick $z-1$ distributions $\{\text{Q}^{\to, (t)}_{j_1}, ...,
\text{Q}^{\to, (t)}_{j_{z-1}}\}$ in the current population (one can
exclude the already chosen $\text{Q}^{\to, (t)}_i$, but this is not
necessary), and assign to each of them one of the directions
$\hat{n}_j$ chosen at step (2).

(4) Generate a new field $\text{Q}^{\to, \mathrm{new}}_i$ according to
\refeq{eq:cavity_eq}: 
\newline
$\text{Q}^{\to, \mathrm{new}}_i = \mathcal{F}_{\to}(\{\text{Q}^{\to, (t)}_{j_1}, ...,
\text{Q}^{\to, (t)}_{j_{z-1}}\})$
and assign it to $\text{Q}^{\to, (t)}_i$

(5) Repeat operations (1) to (4) $m$ times, and increment the
iteration time by $1$. Now, rename the population as
$\{\text{Q}^{\to, (t+1)}_1, ..., \text{Q}^{\to, (t+1)}_{m} \}$,

(6) Test the convergence of the obtained $\mathcal{Q}(\text{Q}^{\to})$. Again, due
to the complexity of the object $\mathcal{Q}(\text{Q}^{\to})$, testing
the convergence is hard to do numerically, as it requires a very large
population of $\{\text{Q}^{\to}\}$ to describe the $\text{Q}^{\to}$
space precisely enough. However, in this work we will use the
Population Dynamics only for determining the satisfiability threshold
$\zcmm$. Hence for our purpose in this work, we do not need to
describe $\mathcal{Q}(\text{Q}^{\to})$ in detail, since we just need
to know if it exists. We thus adopt a simpler criterion to test the
convergence of our Population Dynamics. It is based on the convergence
of the average width of the distribution $\{\text{Q}^{\to}\}$. If this
width converges to a finite value, a solution to the cavity equations
exists (satisfiability), whereas if it vanishes, no solution exists
(unsatisfiability). If the width
have converged, we stop here, otherwise, we restart from step (1) for a new
iteration.

One last technical difficulty is to perform the integration involved
in \refeq{eq:cavity_eq} during step (4). The integrand contains a
delta-function via the $\chi_a$ constraint, coming from force and
torque balance constraints. This means the integrand
is non-zero only on a set that has a dimension smaller than that of
the integration domain. A naive integration scheme would therefore
never probe the integrand in this region. The solution to this problem
is a simple change of variable. Let us first rewrite the
$\delta$-functions in the $\chi_a$ appearing in \refeq{eq:cavity_eq} in a linear algebra form:
\begin{equation}\label{eq:interaction_node}
  \begin{array}{rl}
   \chi_a(\{ f^{\,n},f^{\,t},\hat{n}^{\,a},\hat{t}^{\,a} \}_{\partial a}) 
 \hskip-0.1in &  =   \displaystyle \delta
  \Big( \mathbf{b}_i + {\bf A} \mathbf{f}_{\{j\in \partial a-i\}}\Big)\\
  &\qquad\times \displaystyle\prod_{j \in \partial a} \Theta(f_{j}^{\,n}) \Theta(\mu
  f_{j}^{\,n} - f_{j}^{\,t})
  \end{array}
\end{equation}
where $\mathbf{f}_{\{j\in \partial a-i\}}$ is a $2(z-1)$ vector
concatenating the $z-1$ normal and $z-1$ tangential forces, $f^{\,n}_j$ and
$f^{\,t}_j$, for $j\neq i$, $\mathbf{b}_i = (\vec{f}^{\,a}_i,
\vec{r}^{\,a}_i \times \vec{f}^{\,a}_i)$ is a $2D$ vector, and
$\mathbf{A}$ is a $2D$-by-$2(z-1)$ matrix encoding the geometrical
configuration determined by the force directions $\{\hat n_j^{\,a}\}$ and $\{\hat t_j^{\,a}\}$.

The constraint appearing in the $\delta$-function of
\refeq{eq:interaction_node} is now expressed as linear algebra
problem. Now, if $z=D+1$, this linear problem is readily invertible,
but if $z>D+1$ (which is the most common situation), it is an underdetermined problem. The vector $\mathbf{b}_i$ is given, and we
wish to integrate over the space of solutions
$\mathbf{f}_{\{j\in \partial a-i\}}$ of this underdetemined linear problem.
Finding one of the infinitely many solutions to $-\mathbf{b}_i ={\bf
  A} \mathbf{f}_{\{j\in \partial a-i\}}$ is done with a QR
decomposition~\cite{QRd}. The QR decomposition is the fact that the $2(z-1)$-by-$2D$
rectangular matrix ${\bf A}^T$ (transpose of ${\bf A}$) can be
factorized as:
\begin{equation}
 {\bf A}^T = {\bf Q} {\bf R} 
\end{equation}
where ${\bf Q}$ is an $2(z-1)$-by-$2(z-1)$ square orthogonal
matrix, and ${\bf R}$ is  $2(z-1)$-by-$2D$ matrix whose top
$2D$-by-$2D$ part ${\bf R_1}$ is an upper triangular matrix and bottom part is
identically zero. It is easy to verify that a solution to the linear
problem $-\mathbf{b}_i ={\bf
  A} \mathbf{f}_{\{j\in \partial a-i\}}$ is given by:
\begin{equation}
    \mathbf{f}_{\{j\in \partial a-i\}}(\mathbf{c})= {\bf Q}
       \left[\begin{array}{c} ({\bf R_1}^T)^{-1} \mathbf{b}_i\\ \mathbf{c} \end{array}\right] \\ 
\end{equation}
where $\mathbf{c}$ is \emph{any} $2(z-1)-2D$ vector. This freedom of
choice for $\mathbf{c}$ corresponds exactly to the $2(z-1)-2D$
dimensional space of solution. So, eventually, the cavity equation \refeq{eq:cavity_eq} can be
rewritten as:
\begin{equation}
  \begin{array}{rl}
  & \text{Q}^{a\to i}(f_i^{\,n},f_i^{\,t}) =  \displaystyle \frac{\Theta(f_{i}^{\,n}(\mathbf{c})) \Theta(\mu
  f_{i}^{\,n}(\mathbf{c}) - f_{i}^{\,t}(\mathbf{c}))}{Z^{a \to
        i}} \displaystyle \int \dd \mathbf{c} \dd
    \hat{t}_i^{\,a} \! \!\\
   & \hskip-0.2in \times \!\!\! \displaystyle \prod_{\substack{j\in \partial a - i \\c  = \partial
        j - a }}   \! \! \dd \hat{t}_j^{\,a}\, \text{Q}^{c\to
      j}(f^{\,n}_j(\mathbf{c}), f^{\,t}_j(\mathbf{c})) \Theta(f_{j}^{\,n}(\mathbf{c})) \Theta(\mu
  f_{j}^{\,n}(\mathbf{c}) - f_{j}^{\,t}(\mathbf{c}))
  \end{array}
  \label{eq:3}
\end{equation}
with a Jacobian of $1$.

\subsection{Force distribution at isostaticity}

Obtaining a solution to \refeq{eq:cavity_averaged} via the Population Dynamics
algorithm as described above is \emph{a priori}
enough to get the force distribution through
\refeq{fdis}. Unfortunately, in practice the sampling provided by the
population used in the Population Dynamics is too small to get a
detailed description of the force distribution.

Quite fortunately however, drastic simplifications of the
self-consistent cavity equations \refeq{eq:cavity_eq} and \refeq{eq:cavity_averaged} occur 
at the isostatic point which allow us to access the force distribution in an easier way. Indeed, in an isostatic configuration, there exists a unique set of
solutions of the contact forces for a given graph (at a given
pressure), meaning that the
force probabilities $\text{Q}^i(f^{\,n},f^{\,t})$ on every contact are
$\delta$-functions. 
This, in turn, constraints the distributions
$\text{Q}^{a\to i}(f^{\,n},f^{\,t})=\text{Q}^{b\to
  i}(f^{\,n},f^{\,t})=\text{Q}^i(f^{\,n},f^{\,t})$ to be also $\delta$-functions. This
trivializes the cavity equations \refeq{eq:cavity_eq} to:
\begin{equation}
  \label{eq:trivial}
  \begin{array}{ll}
    & \hskip-0.2in \text{Q}^{a\to i}(f_i^{\,n},f_i^{\,t})   = \chi_a(\{ f^{\,n},f^{\,t},\hat{n}^{\,a},\hat{t}^{\,a} \}_{\partial a}) = \delta
  \Big( \vec{f}^{\,a}_i + \!\!\! \displaystyle \sum_{ \in \partial a-i}  \vec{f}^{\,a}_j \Big)\\
 & \qquad \times \delta
  \Big( \displaystyle \vec{r}^{\,a}_i \times \vec{f}^{\,a}_i+  \!\!\!  \!\!\!  \sum_{j
    \in \partial a-i}  \vec{r}^{\,a}_j \times \vec{f}^{\,a}_j  \Big)
  \Theta(f_{i}^{\,n}) \: \Theta(\mu  f_{i}^{\,n} - f_{i}^{\,t}).
  \end{array}
\end{equation}

Using \refeq{eq:cavity_averaged} and \refeq{fdis}, we then obtain that the overall
force distribution in an isostatic configuration satisfies:
\begin{equation}
 \begin{array}{rl}
\text{P} (f^{\,n},f^{\,t}) & \!\!\! \propto \displaystyle \sum_{z} z \, \text{R}(z) \!\! \displaystyle \int \Omega(\hat{n}_i,\{\hat{n}_j\})\!
  \dd \hat{n}_i \dd \hat{t}_i \\
  & \hskip-0.5in \times \displaystyle \prod_{j=1}^{z-1} \left[\dd \hat{n}_j\dd \hat{t}_j \dd f_j^{\,n} \dd f_j^{\,t}
      \, \text{P} (f_j^{\,n},f_j^{\,t} )\right]  \chi_a(\{ f^{\,n},f^{\,t},\hat{n},\hat{t} \})
  \end{array}
\label{eq:P_of_F_at_transition}
\end{equation}
where the proportionality constant is just the normalization.
This is a self-consistent equation of the form $\text{P} (f^{\,n},f^{\,t}) =
\mathcal{L}[\text{P} (f^{\,n},f^{\,t})]$ that can be solved iteratively. Starting
with an initial distribution (`guess') $\text{P}^{(0)} (f_j^{\,n},f_j^{\,t}
)$, we iterate through $\text{P}^{(i+1)} (f^{\,n},f^{\,t}) =
\mathcal{L}[\text{P}^{(i)} (f^{\,n},f^{\,t})]$, and obtain the force
distribution via $\text{P} (f^{\,n},f^{\,t})
= \text{P}^{(i\to \infty)}(f^{\,n},f^{\,t})$. In practice, a few tens of iterations are
sufficient to obtain convergence.

\section{Results}

\subsection{Force distribution for frictionless spheres packings}
\label{sec:pofF}

\begin{figure}[!]
\centering
    \includegraphics[width=0.48\textwidth]{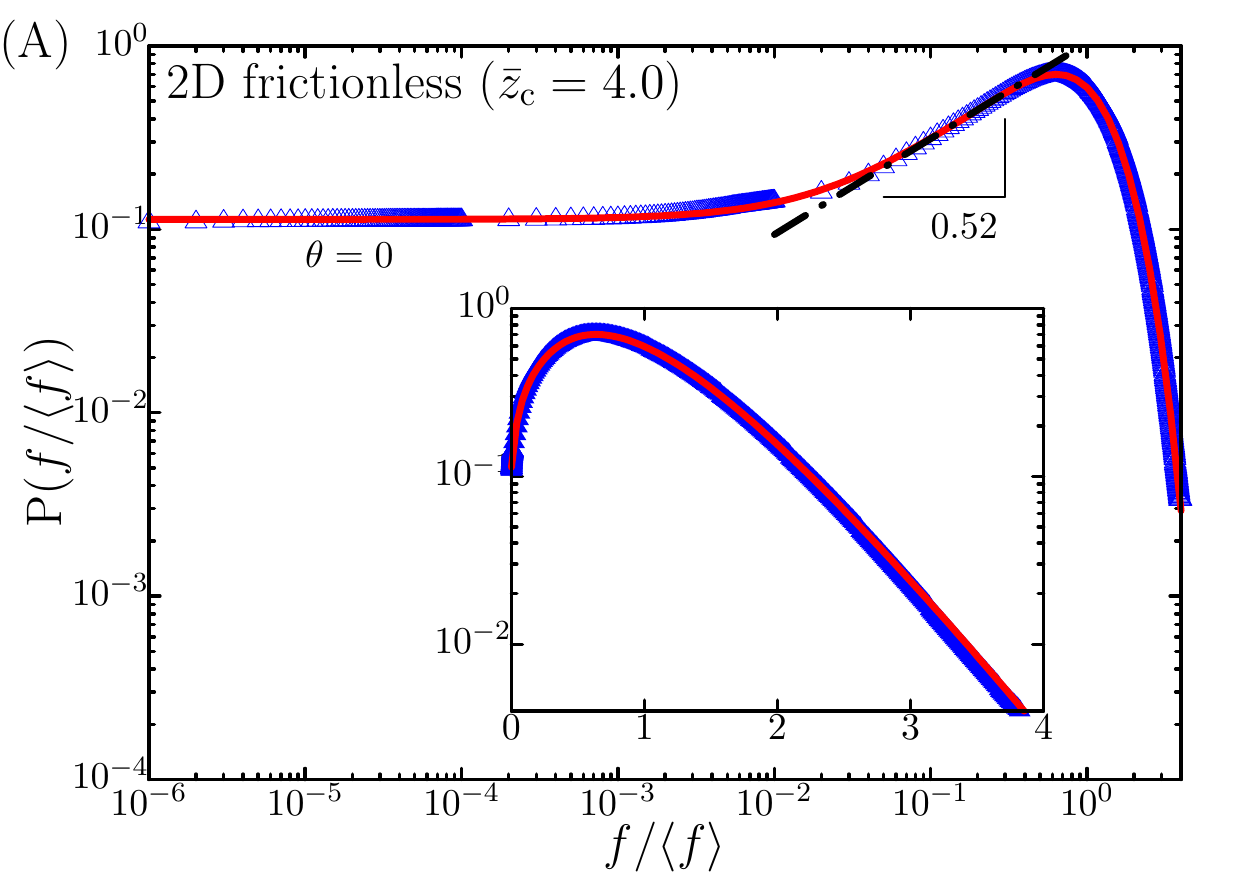}
    \includegraphics[width=0.48\textwidth]{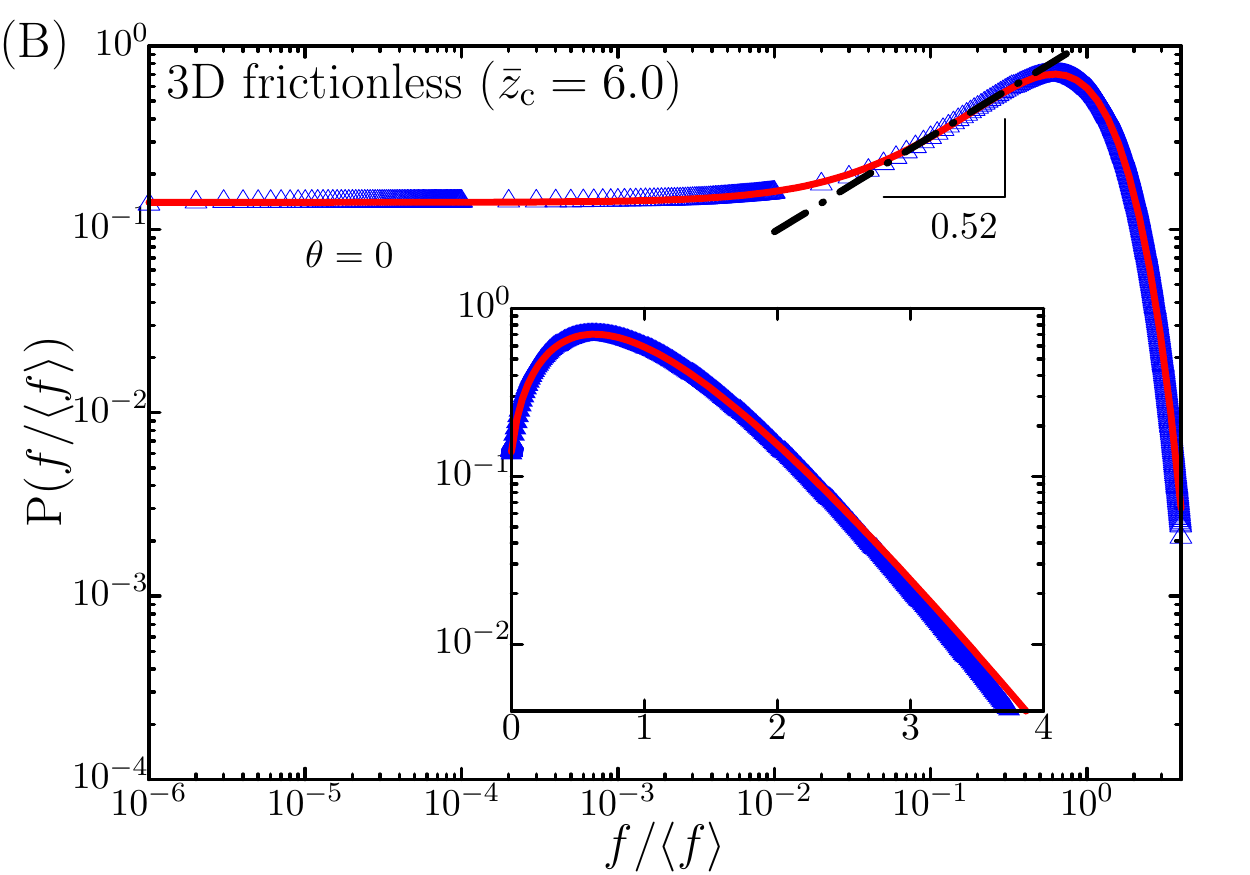}
    \begin{minipage}{0.5\textwidth} 
    \includegraphics[width=0.96\textwidth]{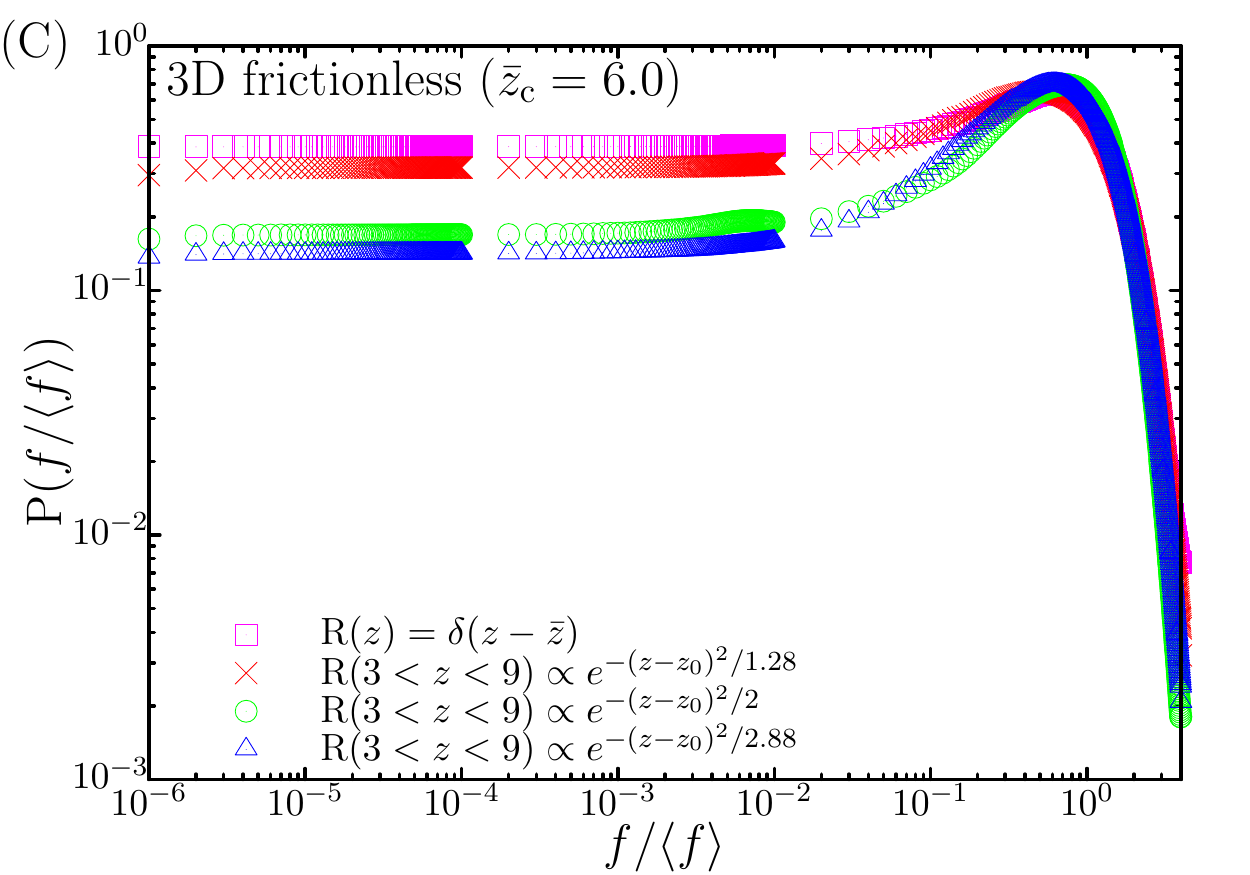}
  \end{minipage}%
  \begin{minipage}{0.5\textwidth}
   \caption{Force distribution, $\text{P}(f)$ in frictionless spheres
     packing in (A) 2-dimensional and (B) 3-dimensional systems.
     Results obtained from the cavity method (open triangles) show a
     flat regime (in a log-log plot) with exponent $\theta= 0$ at small forces and $0.52$
     in the intermediate region in both cases. We use here as a
     contact number distribution $\text{R}(3<z<9) \propto
     e^{-(z-z_0)^2/2.88}$. In inset, log-linear plot of the same
     distribution exhibits approximate exponential tail at large
     forces. The red solid lines correspond to the fitting functions $P(x)$,
     as defined in the text. (C) Comparison of force distributions
     obtained with narrower distributions $\text{R}(z)$ (at fixed average $\bar{z}$). The exponent $\theta= 0$ is obtained independently of the width of $\text{R}(z)$, but the peak in the force distribution becomes smaller when using a narrower $\text{R}(z)$.}\label{fig:pf}
  \end{minipage}
\end{figure}

We start by computing the force distribution ${\text P}(f=f^{\,n})$ for two and three-dimensional frictionless
spheres packings, when we fix the average contact number ${\bar z}=\zc=4.0$ and $6.0$
respectively, by solving~\refeq{eq:P_of_F_at_transition}. Results in \reffig{fig:pf} reproduce the force distributions similar as seen in numerical simulations~\cite{Radjai1996,
  OHern2001,PhysRevE.62.2510,Makse2000,Silbert2002,PhysRevE.72.011301,Donev2005,lerner2013low}
and experiments~\cite{Liu28071995, PhysRevE.57.3164, PhysRevE.60.5872,Erikson2002,Makse2000,Brujic2003, Brujic2003b}.  The force distribution we obtain can be well fitted with $P(x)=[7.84x^2+0.86-0.75/(1+4.10x)]e^{-2.67x}$ for 2-D (\reffig{fig:pf}A) and $P(x)=[7.45x^2+1.20-1.06/(1+2.33x)]e^{-2.65x}$ for 3-D (\reffig{fig:pf}B), with $x=f/\langle f \rangle$. Both fitting functions are close to the empirical fit $P(x)=[3.43x^2+1.45-1.18/(1+4.71x)]e^{-2.25x}$ to the force distribution of dense amorphous packings generated by Lubachevsky-Stillinger algorithm in 3-D by Donev {\it et al.}~\cite{Donev2005}. Note that although the tails of the force distributions can be well fitted to exponential, claims of the precise form of the tails are difficult to conclude, as the presented data only varies over $2$ decades. 

Our method allows to access the small force region with unprecedented
definition. We gather data down to $10^{-6}$ times the peak force
(which is of the order of the pressure). This range is way below what
is accessible with state of the art simulations of packings
\cite{Radjai1996, OHern2001,PhysRevE.62.2510,PhysRevLett.109.205501,
  Silbert2002,PhysRevE.72.011301,lerner2013low}.
The reason for this is that we avoid two problems: (i) we work
with~\refeq{eq:P_of_F_at_transition} directly in the thermodynamic
limit, whereas simulated packings typically are limited to few tens of
thousands of forces~\cite{Radjai1996,
  OHern2001,PhysRevE.62.2510,Silbert2002} which limits the definition
of the obtained force distribution, and (ii) we can work exactly at
the jamming transition point, as we set ${\bar z}=2D$, which contrasts
with actual numerical or experimental studies where the limit of
vanishing pressure with ${\bar z}=2D$ is very challenging.

The behavior of $\text P(f)$ at small forces has recently attracted
attention
\cite{PhysRevLett.109.125502,PhysRevLett.109.205501,lerner2013low,degiuli_distribution_2014},
due to its central role for the mechanical stability of
packings~\cite{PhysRevLett.109.125502,lerner2013low}. 
Lerner et al.~\cite{lerner2013low} pointed out a relation between the small force
scaling $\text P(f) \sim f^{\theta}$ and the distribution function of
the gaps $h$ between particles close to contact $g(h)\sim h^{-\gamma}$ via the inequality
$\gamma \geq (1-\theta)/2$. 
Few empirical data exist so
far for the $\theta$ exponent. Some recent efforts greatly
improved the data statistics~\cite{PhysRevLett.109.205501,lerner2013low,degiuli_distribution_2014},
but however still widely disagree over the value of the exponent, which is
found anywhere between $\theta\approx 0.2$ and $\theta\approx
0.5$. This calls for more insight from theory.

Here we find for frictionless spheres a distribution of contact forces
having a finite value for $f=0$ in the mean-field approximation. This
translates as an exponent $\theta = 0$ over four decades of data in
both 2-D and 3-D packings (\reffig{fig:pf}). 
We stress here that this observation is not dependent
on the input distribution of coordination number $\text{R}(z)$. The comparison of the force distributions we obtain with various shapes of $\text{R}(z)$ is shown in \reffig{fig:pf}C. In particular, the exponent $\theta=0$ as well as the large-force tail stays the same when we
change $\text{R}(z)$ from the empirical fit $\text{R}(3<z<9) \propto e^{-(z-z_0)^2/2.88}$ (at fixed average $\bar{z}$) to  \textit{e.g.} a regular graph $\text{R}(z) = \delta(z-\bar{z})$ in 3-D packings. 
However, the value of the exponent found in recent
investigations~\cite{PhysRevLett.109.205501,lerner2013low}
seems to be incompatible with the value of $\theta$ obtained from our
theory. The exponent might be dependent on the protocol by which
jammed packings are generated~\cite{lerner2013low}, although this
point is still debated~\cite{degiuli_distribution_2014}. The value
$\theta = 0$ we obtain is an ensemble averaged, mean-field
exponent. The value of the mean-field exponent might be lost in a
finite-dimensional packing, but the fact that the exponent seems to be
dimension independent~\cite{PhysRevLett.109.205501}
is an encouraging sign for a mean-field approach, as the mean-field
value should be valid in high dimension. We note that a very recent
work by Charbonneau et al. finds an exponent $\theta \approx 0.42$ in
infinite dimensions from replica
theory~\cite{kurchan_exact_2013,charbonneau_exact_2013,charbonneau_fractal_2014}.
Our exponent is incompatible with
the relations $\gamma \geq 1/(2+\theta)$ derived by
Wyart~\cite{PhysRevLett.109.125502}, or $\gamma \geq (1-\theta)/2$
derived by Lerner et al~\cite{lerner2013low}, where $\gamma$ is the
exponent describing the distribution function of the gaps $h$ between
particles close to contact $g(h)\sim h^{-\gamma}$. Those two different
relations are obtained from considerations of mechanical stability
against respectively extended or local (buckling) excitations. In 3-D,
the local excitations seem to be the dominant ones, and thus only the
relation $\gamma \geq (1-\theta)/2$ should hold~\cite{lerner2013low}
in that case. In our mean-field framework, we miss the relation
between excitation modes and force distribution because we neglect the
spatial structure of finite dimensional packings. In fact, the pair
correlation function in our framework has no structure beyond
contact. Particle positions are omitted, apart from a local constraint
of non overlaping, and there is therefore no excitation mode in our
approach.

On the theoretical side, several values of $\theta$ have been
predicted: the replica theory at the 1-RSB level also predicts a finite value for $\text P(f=0)$~\cite{RevModPhys.82.789,berthier2011microscopic,PhysRevLett.109.205501}
(it predicts a Gaussian for $\text P(f)$), but the fullRSB solution
gives $\theta \approx 0.42$ in infinite
dimensions~\cite{kurchan_exact_2013,charbonneau_exact_2013,charbonneau_fractal_2014}; the q-model can give several
values for $\theta$, depending on the underlying assumptions, but in
any case predicts $\theta\geq 1$~\cite{Liu28071995,PhysRevE.53.4673}; 
and Edwards' model predicts $\theta=1/({\bar z}-2)$~\cite{Brujic2003,
  Brujic2003b}. The differences
between Edwards' and q-model approaches, which mostly stem from the
treatment of local disorder in the contact normal directions, indicate
that the way to deal with this disorder is crucial for a correct
description of the small force behavior.

Interestingly, we find an intermediate regime for $f$ slightly smaller
than its average value $\langle f \rangle$, for which $\text P(f) \sim
f^{0.52}$ (\reffig{fig:pf}A, \reffig{fig:pf}B). We observe that this pseudo-exponent gets smaller when using a narrower $\text{R}(z)$ as input (\reffig{fig:pf}C). The smallest value it
can reach for regular graph $\text{R}(z) = \delta(z-\bar{z})$ are
$\sim0.40$ in 2-D and $\sim0.21$ in 3-D packings respectively. 
The results of this regime, extending from $f \simeq 10^{-2}\langle f \rangle$ to $f \simeq
\langle f \rangle$ is the one probed by most experiments and
simulations. This suggests that careful measurements at very small $f$
are needed to avoid pre-asymptotic behavior in the estimation of
$\theta$ from experimental or numerical data, which might explain a
lot of discrepancies observed in the current literature. Only few very
recent numerical data managed to reach this
regime~\cite{lerner2013low,degiuli_distribution_2014}.


\subsection{Calculation of $\zcmm$ for sphere packings with arbitrary friction coefficient}

\label{sec:zcm}
\begin{figure}[htb]
\centering
    \includegraphics[width=0.48\textwidth]{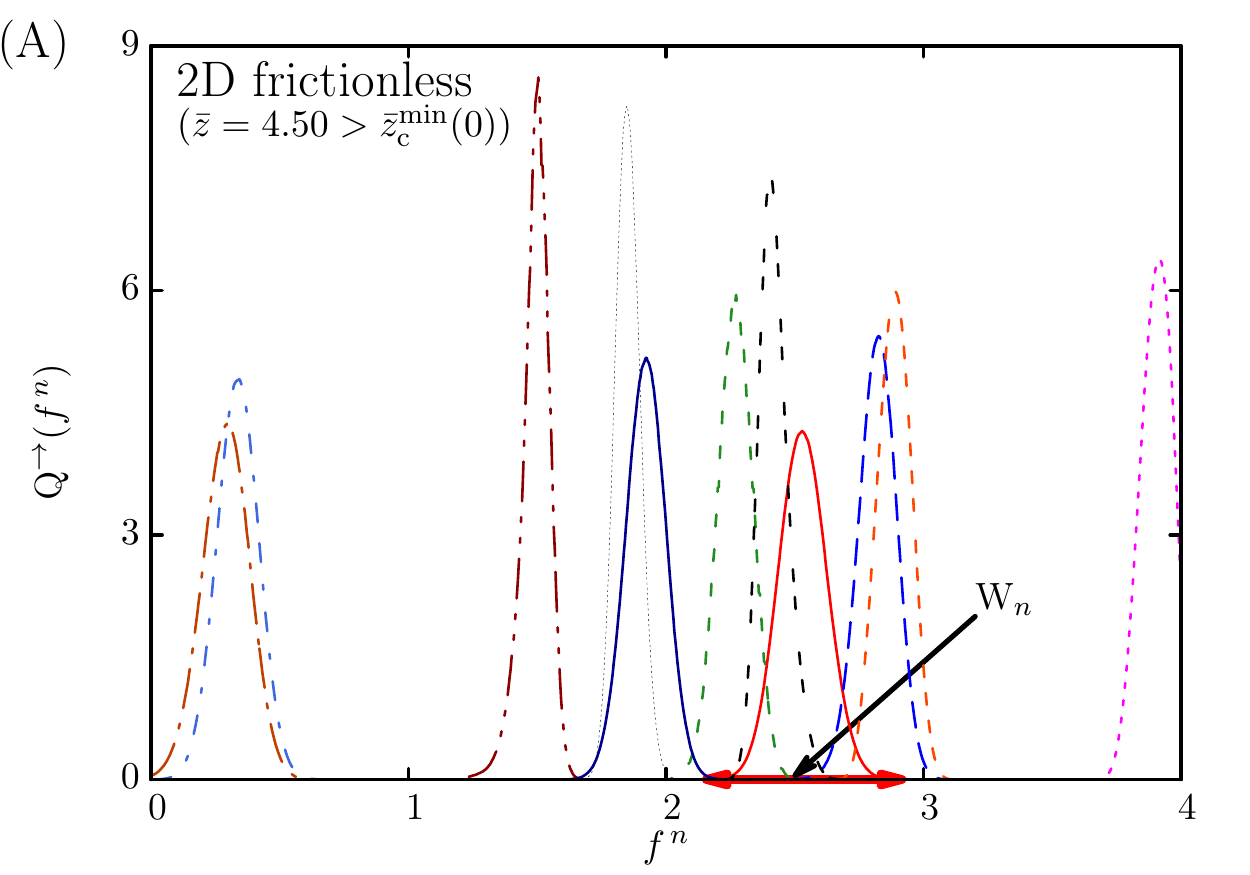}    
    \includegraphics[width=0.48\textwidth]{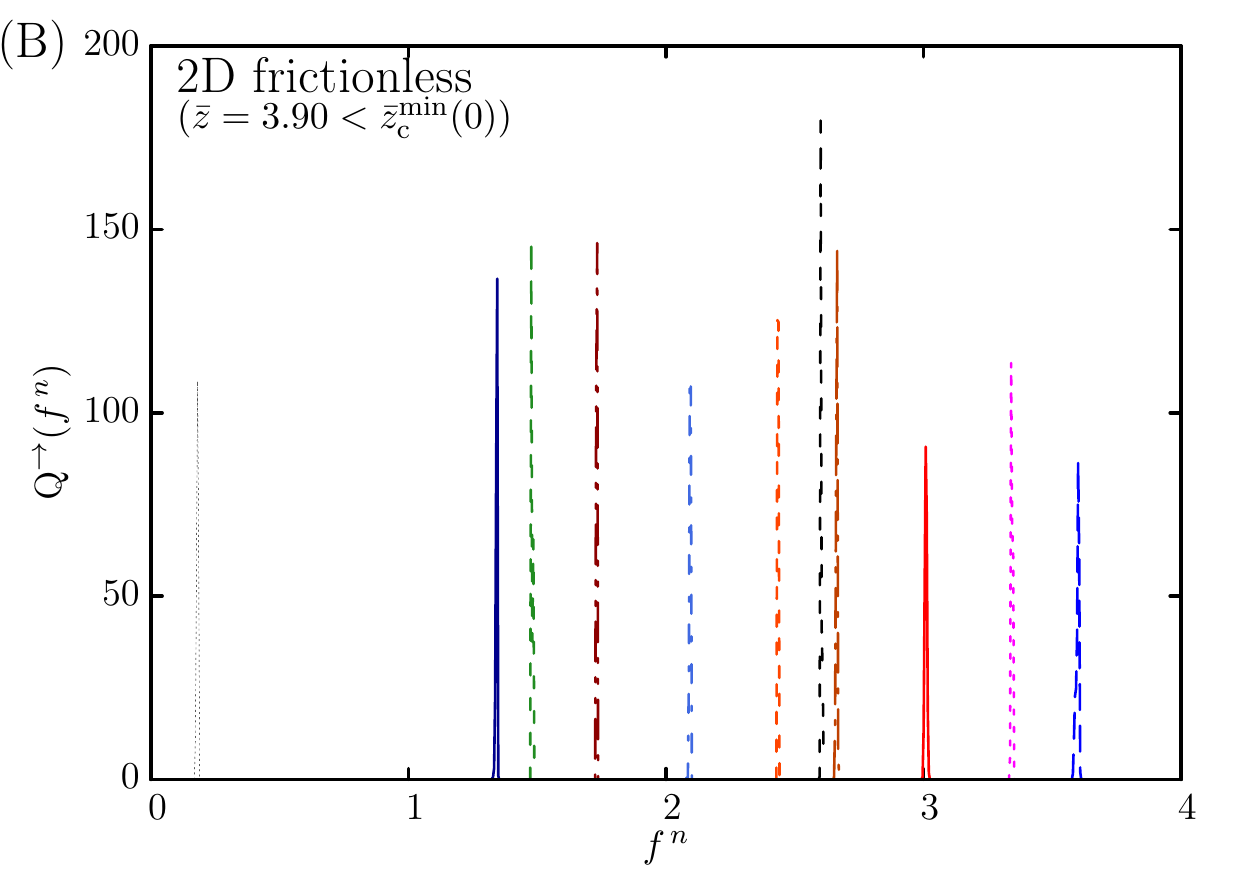}    
  \caption{
 The force distribution $\{\text{Q}^{\to}(f^{\,n})\}$ when (A) $\bar{z} = 4.50 > \zcm(0)$ and (B) when $\bar{z} = 3.90 < \zcm(0)$ for 2-D frictionless spheres packing. In (A), the `width' of $\text{Q}^{\to}(f^{\,n})$ on $f^{\,n}$ 
is defined by $\text{W}_n$ as the difference of two extreme values of $f^{\,n}$ at which $\text{Q}^{\to}(f^{\,n})$ is equal to $10^{-3}$.
  }\label{fig:width}
\end{figure}

We turn to the determination of the force distribution for arbitrary
friction coefficient $\mu$ and a lower bound on $\zc$ for the
existence of random packings of spheres at a given $\mu$. This
threshold corresponds to the point where solutions of the cavity
equations \refeq{eq:cavity_eq} and \refeq{eq:cavity_averaged} no
longer exist. 

We search for the existence of a solution by applying the Population
Dynamics algorithm described in section~\ref{sec:PD}.  A solution
exists if this process leads to a converged averaged width of the
distributions of the population $\{\text{Q}^{\to}\}$ used to describe
$\mathcal{Q}(\text{Q}^{\to})$. Even though this algorithm does not
require a very large population (as we do not seek to obtain a
detailed description of $\mathcal{Q}(\text{Q}^{\to})$ and only want to
know if it exists), its computational cost is still high, and we here
apply the method to frictionless packings and 2-D frictional packings
only.

\begin{figure*}[!t]
\centering
\begin{minipage}{1\textwidth}
\includegraphics[width=0.48\textwidth]{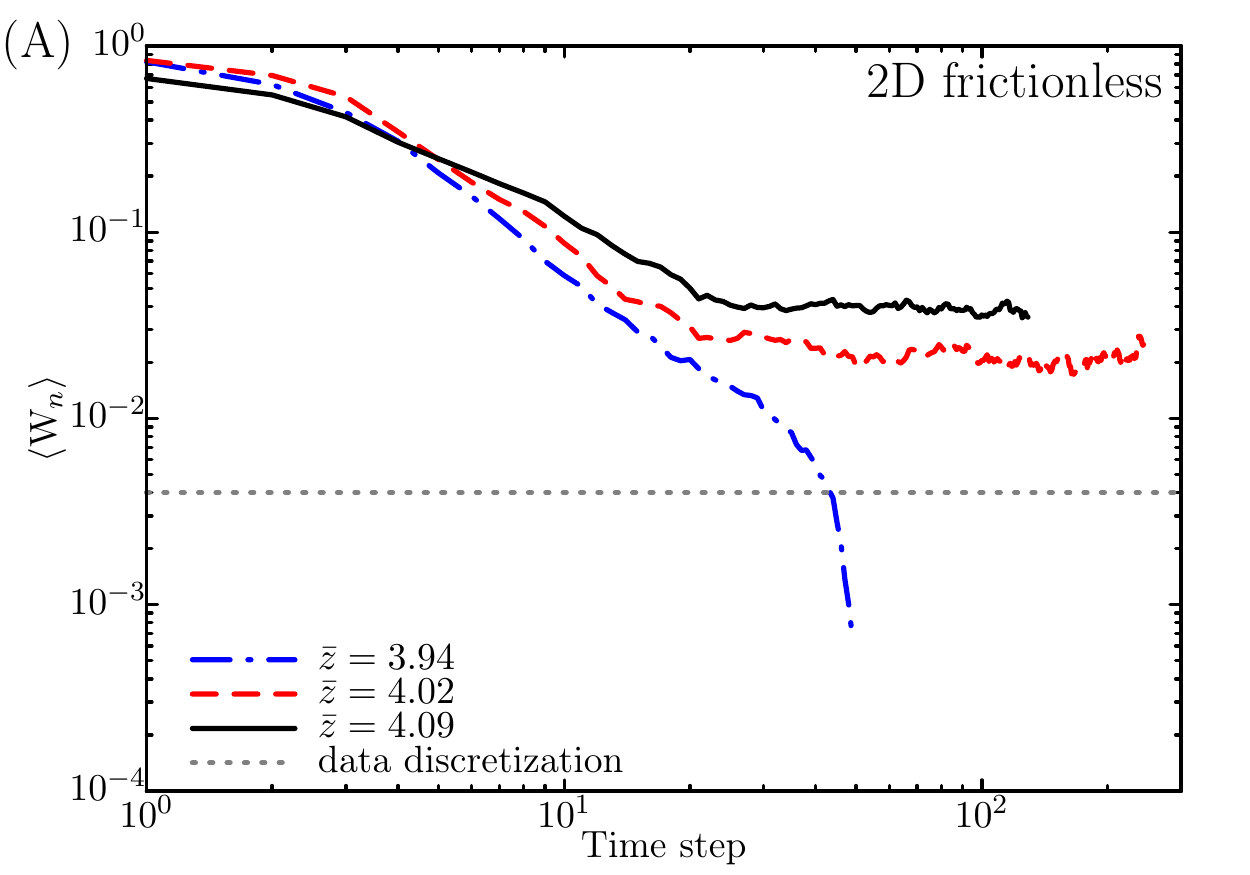}
\includegraphics[width=0.48\textwidth]{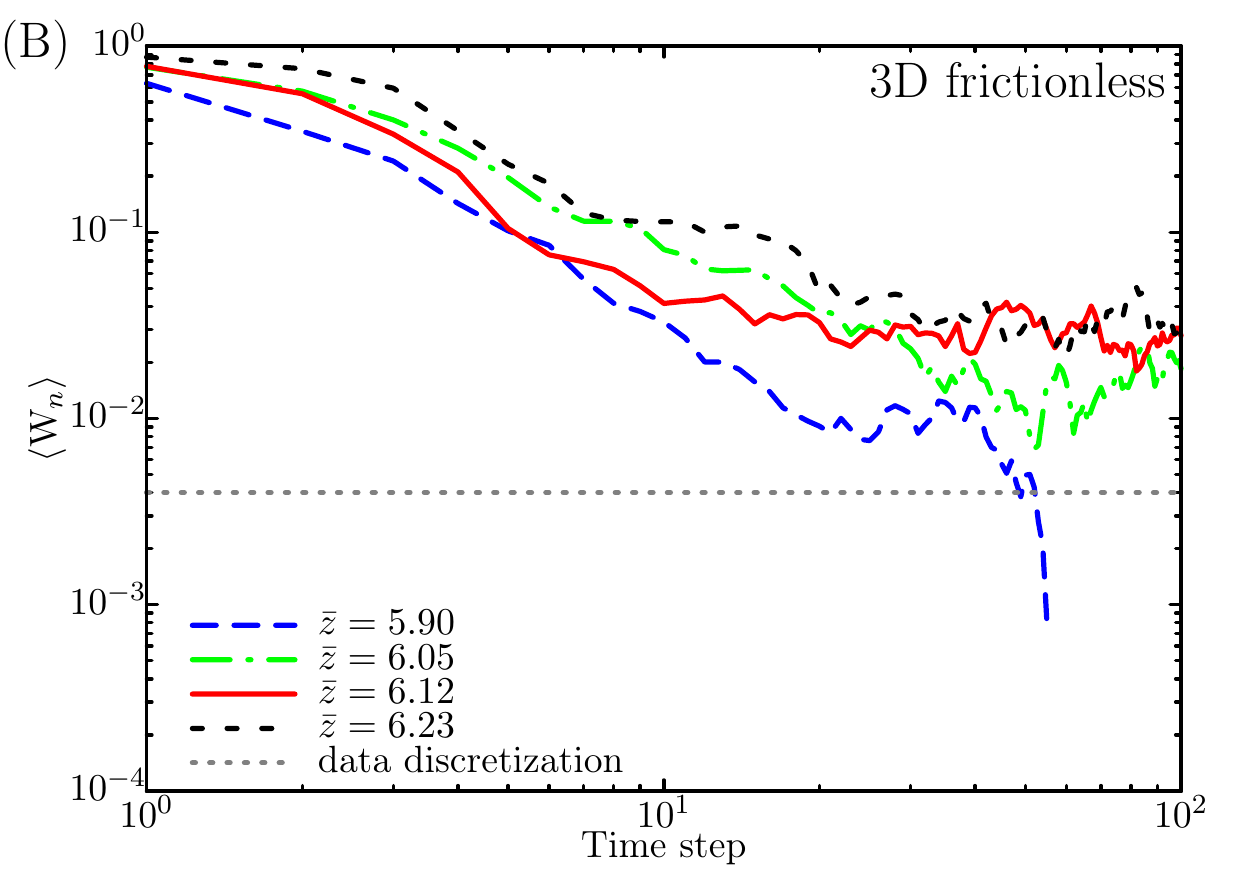} 
 \end{minipage} 
\\
\begin{minipage}{1\textwidth}
\includegraphics[width=0.48\textwidth]{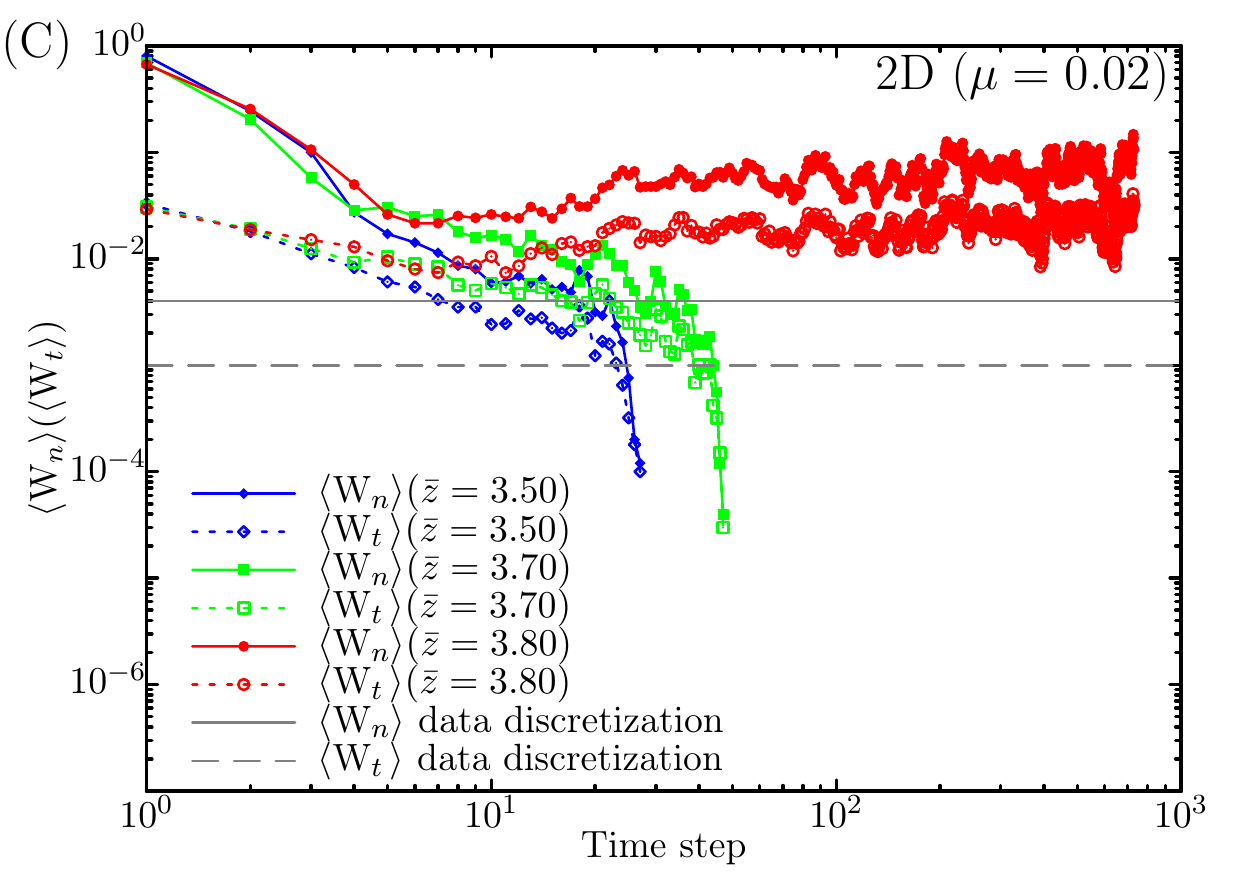}    
 \includegraphics[width=0.48\textwidth]{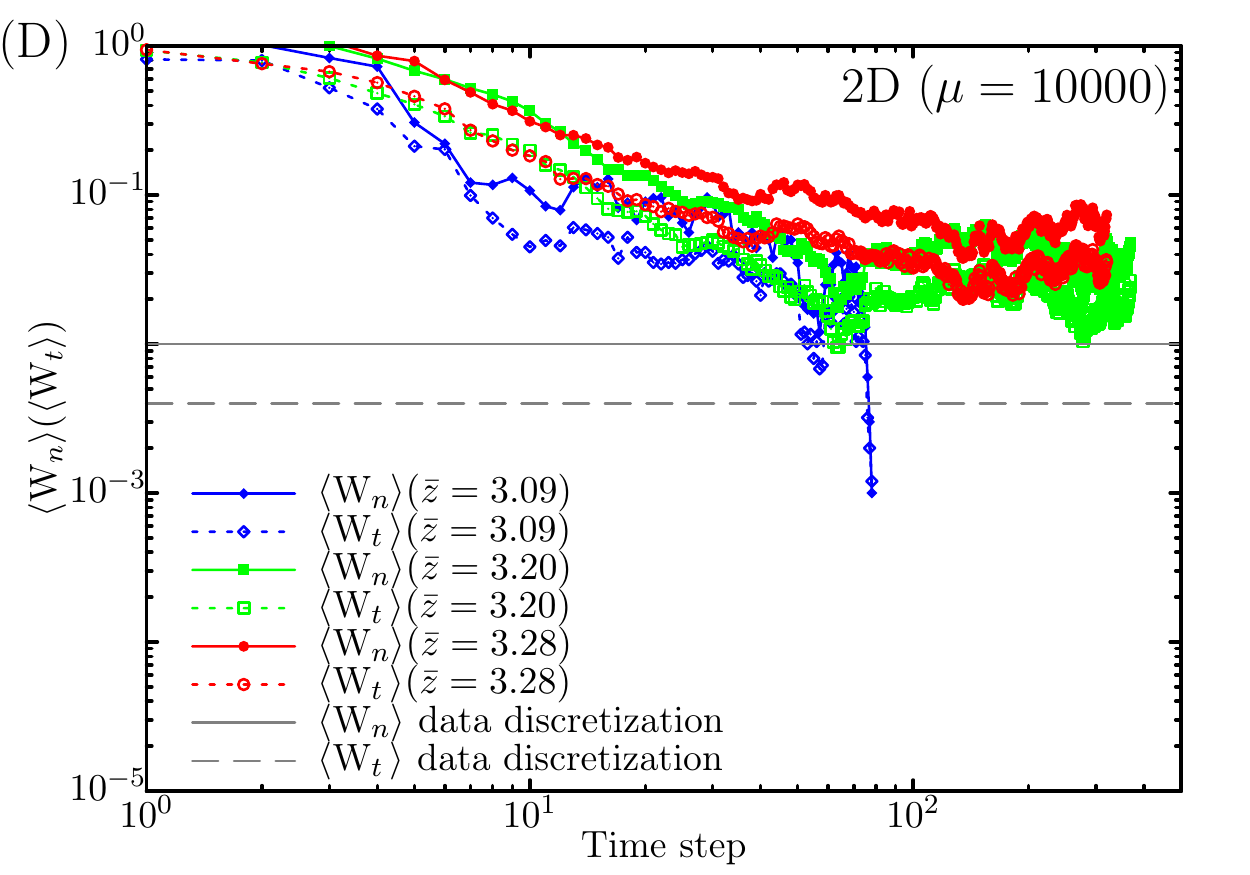}  
\end{minipage}
 \caption{
 (A) Evolution of average width $\langle \text{W}_n \rangle$ (and $\langle \text{W}_t \rangle$ if frictional) of a large population $\{\text{Q}^{\to}\}$ versus time step in population dynamics in (A) 2-D frictionless, (B) 3-D frictionless, (C) 2-D frictional packing with $\mu=0.02$ and (D) 2-D frictional packing with $\mu=10000$. $\zcmm$ is found at (A) $\zcm(\mu=0) \in [3.94,4.02]$ in 2-D frictionless spheres packing, (B) $\zcm(\mu=0) \in [5.90,6.05]$ in 3-D frictionless spheres packing, (C) $\zcm(\mu=0.02) \in [3.70,3.80]$ and (D) $\zcm(\mu=10000) \in [3.09,3.20]$ in 2-D frictional spheres packing.
  }\label{fig:wt}
\end{figure*}

 \begin{figure}[!t]
\centering
    \includegraphics[width=0.48\textwidth]{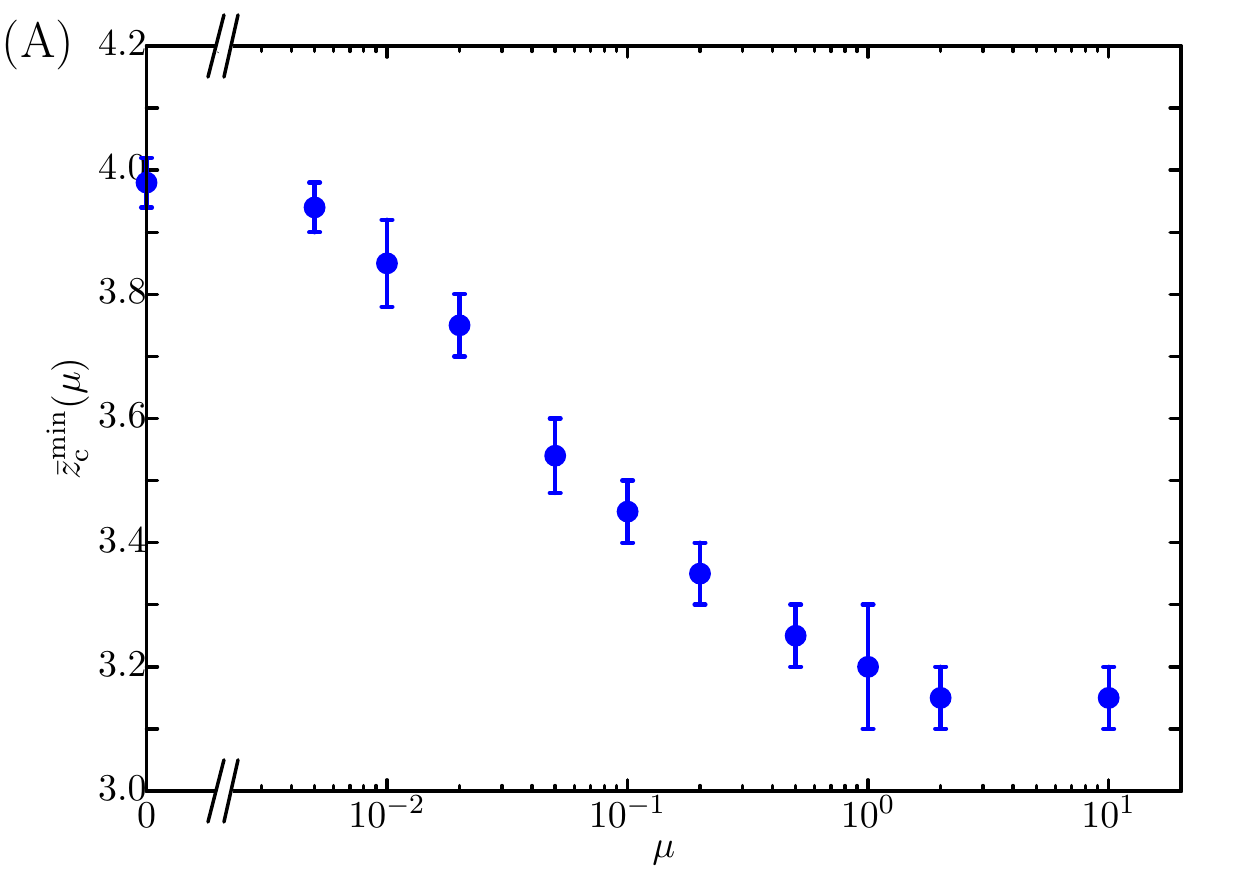}    
    \includegraphics[width=0.48\textwidth]{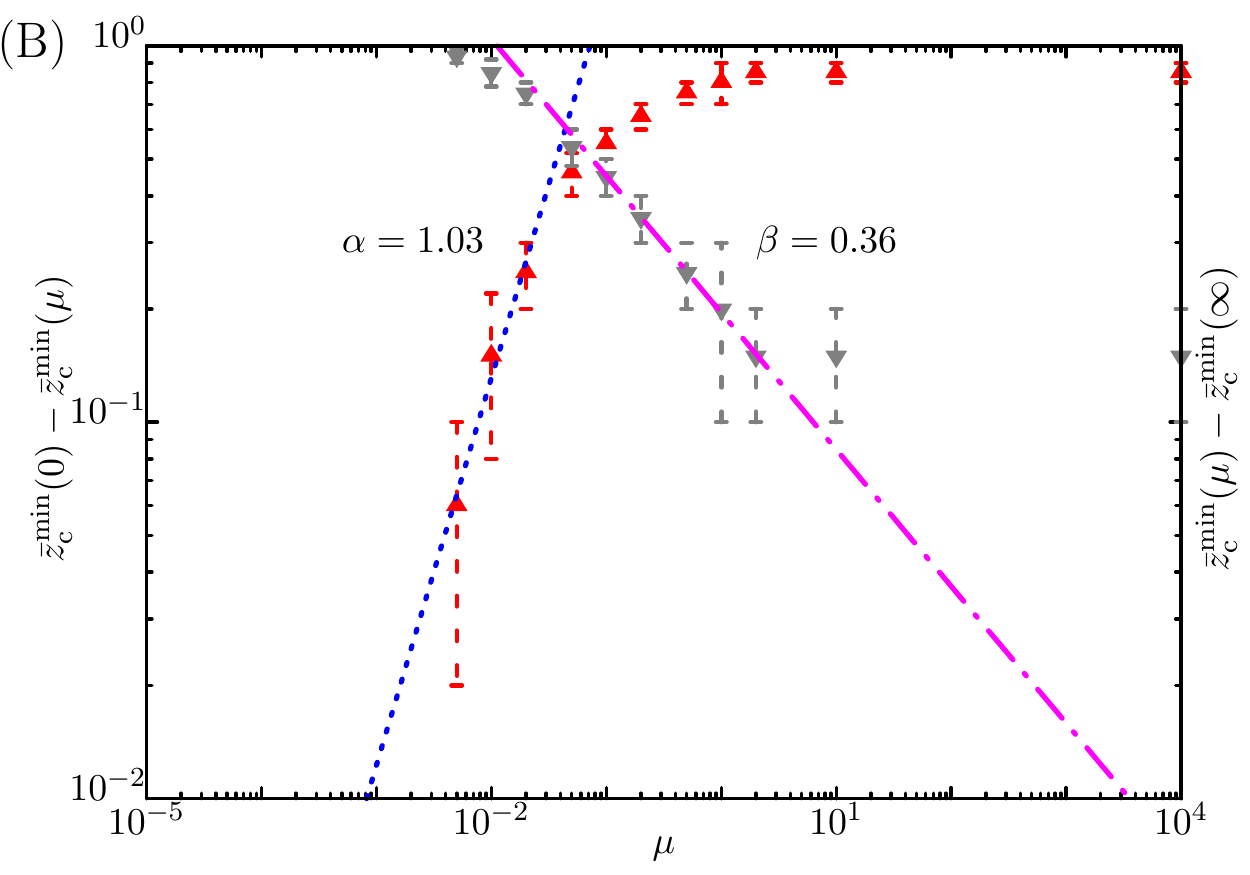}    
  \caption{
(A) Linear-Log plot of the full curve $\zcmm$ vs. $\mu$ for various friction coefficients in 2-D spheres packing.  $\zcmm$ shows a monotonic decrease with increasing $\mu$ from $\zcm({\mu=0})=2 D=4$ to $\zcm({\mu=\infty}) \gtrsim D+1 = 3$. Error bar indicates the range from the largest $\zcmm$ having no solution to the smallest $\zcmm$ having solution. Data point represents the mean of the range. (B) Two power law scaling relations $\zcm(0)-\zcm({\mu}) \sim {\mu}^{\alpha}$, $\zcm({\mu})-\zcm({\infty}) \sim {\mu}^{-\beta}$ are found with $\alpha=1.03$, $\beta=0.36$, respectively.
  }\label{fig:zmu}
\end{figure}

For frictionless packings, the `width' of a force distribution $\text{Q}^{\to}(f^n)$, denoted by $\text{W}_n$, is defined 
as the difference of two extreme values of $f^n$ at which $\text{Q}^{\to}(f^n)$ is equal to $10^{-3}$ (see \reffig{fig:width}A). 
To determine the existence of solution, we calculate the average width over the sites as  $\langle \text{W}_n \rangle$. 
\reffig{fig:wt}A shows the evolution of
the average width of distributions at different $\bar{z}$ versus time
step in population dynamics for the particular case of 2-D
frictionless spheres packings. Results indicate that the final population of distributions $\{ \text{Q}^{\to}\}$
after $t_{\rm max}$ iterations have dramatically different shapes at
various $\bar{z}$. We find that the population $\{\text{Q}^{\to}\}$ rapidly
tends to a set of non-overlapping Dirac peaks when $\bar{z}$ is small
(shown in \reffig{fig:width}B for $\bar{z}=3.90$). In this case, the average width of
fields  decreases as a function of time step in the Population Dynamics iteration and finally goes below data discretization (the
interval of force to integrate the force distribution), leading to no
solution of $\text{Q}^{i}(f^{\,n})$ according to
\refeq{eq:P2Pto}.  Thus the system is overdetermined in the
sense that $\bar{z}$ contacts per particle are not enough to stabilize the
whole packing. In contrast, when $\bar{z}$ is
increased (shown in \reffig{fig:width}A for $\bar{z}=4.50$), the distributions in $\{\text{Q}^{\to}\}$ become broad. The average width converges to a finite
value as seen in~\reffig{fig:wt}A, allowing a non-vanishing force distribution $\text{Q}^{i}(f^{\,n})$ at each contact. 
In this case a set of solutions to force and torque balances exists. 
The range between the largest $\bar{z}$ having no solution and the smallest $\bar{z}$ having solution is the range where $\zcmm$ belongs to. In addition to 2-D frictionless packings, result of 3-D frictionless packings is shown in \reffig{fig:wt}B. The transition points $\zcm(\mu = 0)=2D$ are found in the expected range for both 2-D and 3-D frictionless packings.

For frictional packings, the force distribution $\text{Q}^{\to}(f^{\,n},f^{\,t})$ has the shape of a `blob' that spreads on the $f^{\,n}, f^{\,t}$-plane. In this case the width of the force distribution $\text{W}_n$ and $\text{W}_t$ are defined as the width of the spread from side to side on axis $f^{\,n}$ and $f^{\,t}$ respectively with $\text{Q}^{\to}(f^{\,n},f^{\,t})$ higher than $10^{-3}$: \textit{e.g.} $\text{W}_n = f^{\,n}_\text{max} - f^{\,n}_\text{min}$, where $f^{\,n}_\text{max}$ is the smallest value satisfies the condition $\text{Q}^{\to}(f^{\,n}>f^{\,n}_\text{max},f^{\,t})<10^{-3}$ and $f^{\,n}_\text{min}$ is the largest value satisfies the condition $\text{Q}^{\to}(f^{\,n}<f^{\,n}_\text{min},f^{\,t})<10^{-3}$.
Here we obtain data of average width $\langle \text{W}_n \rangle$ and $\langle \text{W}_t \rangle$ of distributions $\{\text{Q}^{\to}\}$ in 2-D frictional spheres packings with arbitrary friction coefficient. Similarly, the force distributions have different shapes after $t_{\rm max}$ iterations, from which we determine, for example, the satisfiability transition point of force and torque balances in 2-D spheres packing at $\zcm(0.02) \in [3.70,3.80]$ (\reffig{fig:wt}C) and  $\zcm(10000) \in [3.09,3.20]$ (\reffig{fig:wt}D). The full curve $\zcmm$ is shown in \reffig{fig:zmu}A for 2-D frictional spheres packing.  We observe a monotonic
decrease with increasing $\mu$ from $2D=4$ at $\mu=0$, a well-known
behavior of frictional packings, previously found in numerous studies,
both experimentally and numerically
\cite{PhysRevLett.92.054302,PhysRevLett.81.1634,PhysRevE.65.031304,Silbert2002,PhysRevE.72.011301,PhysRevE.75.010301,song2008phase,Wang20103972,C001973A,ohern2013}.
Notice that the critical contact number we obtain at infinite friction
is slightly above the Maxwell argument $D+1$; also a typical
feature~\cite{PhysRevE.65.031304,PhysRevE.75.010301,Wang20103972}. This
is interesting, as it means that the naive counting argument,
ignoring the repulsive nature of the forces, fails to reproduce the
correct bound for such a simple case as sphere packings with $\mu\to
\infty$, where neither Coulomb condition nor non-trivial geometrical
features complexify the picture. The fact that the naive Maxwell counting argument still gives the
correct bound for frictionless sphere packing can therefore be seen as
a quite fortunate isolated prediction.  
In \reffig{fig:zmu}B, two power law scaling relations $\zc^{\rm
  min}(0)-\zcmm \sim {\mu}^{\alpha}, \zcmm - \zcm(\infty) \sim
{\mu}^{-\beta}$ are found with $\alpha=1.03$, $\beta=0.36$ respectively.
Our result $\alpha$ agrees well with the prediction of 2-D monodisperse packing $\alpha=1$ by Wang {\it et al}~\cite{Wang20103972}, and is not far away from previous simulation of 2-D polydisperse packings $\alpha=0.70$~\cite{PhysRevE.75.010301}, while $\beta$ is much smaller than their predicted value $\beta=2$ and the result of $\beta=1.86$ obtained from simulation~\cite{Wang20103972}.

\subsection{Joint force distribution for frictional spheres packings}

\begin{figure*}[!t]
\centering
\begin{minipage}{1\textwidth}
\includegraphics[width=0.3\textwidth]{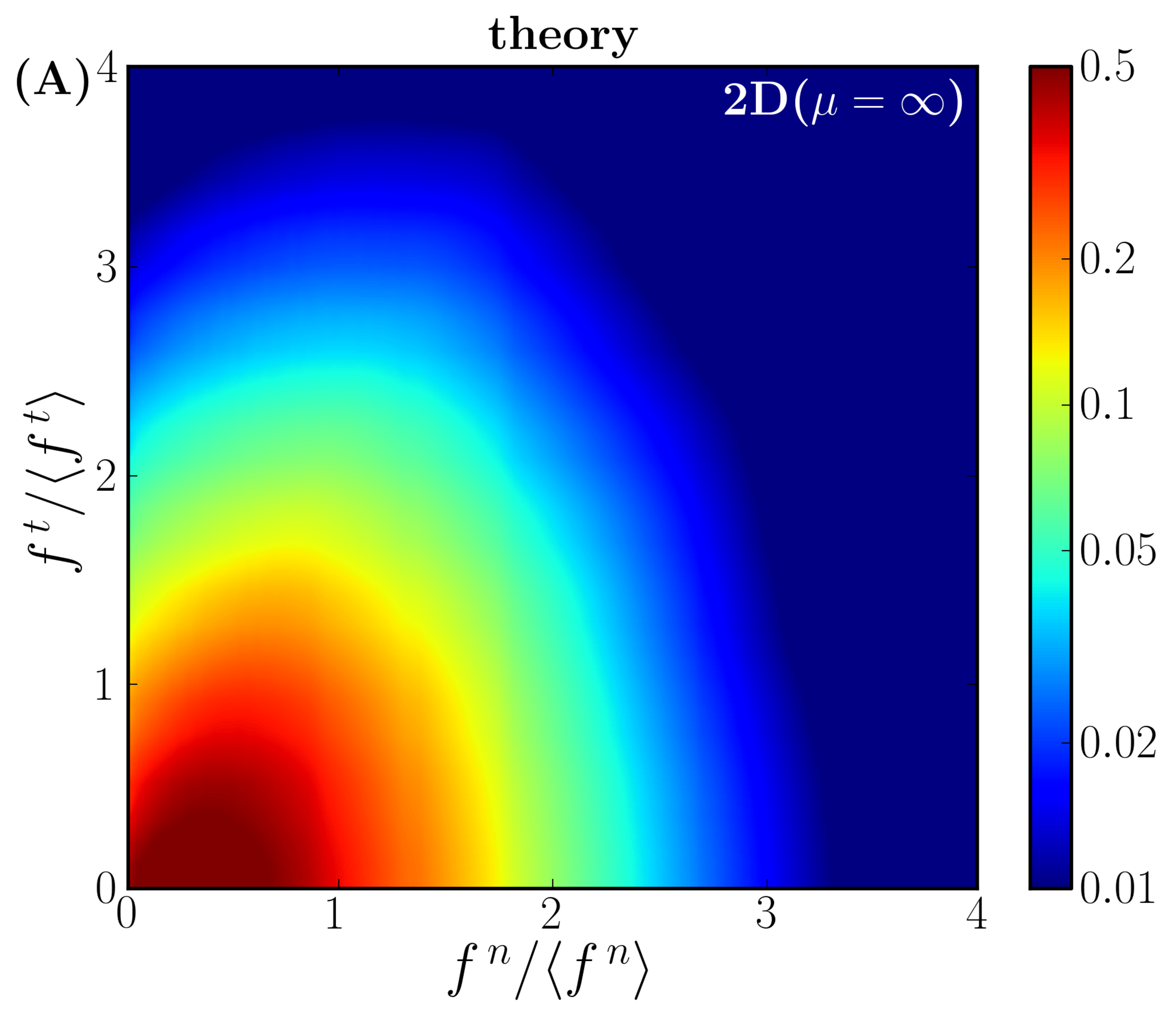}
\includegraphics[width=0.3\textwidth]{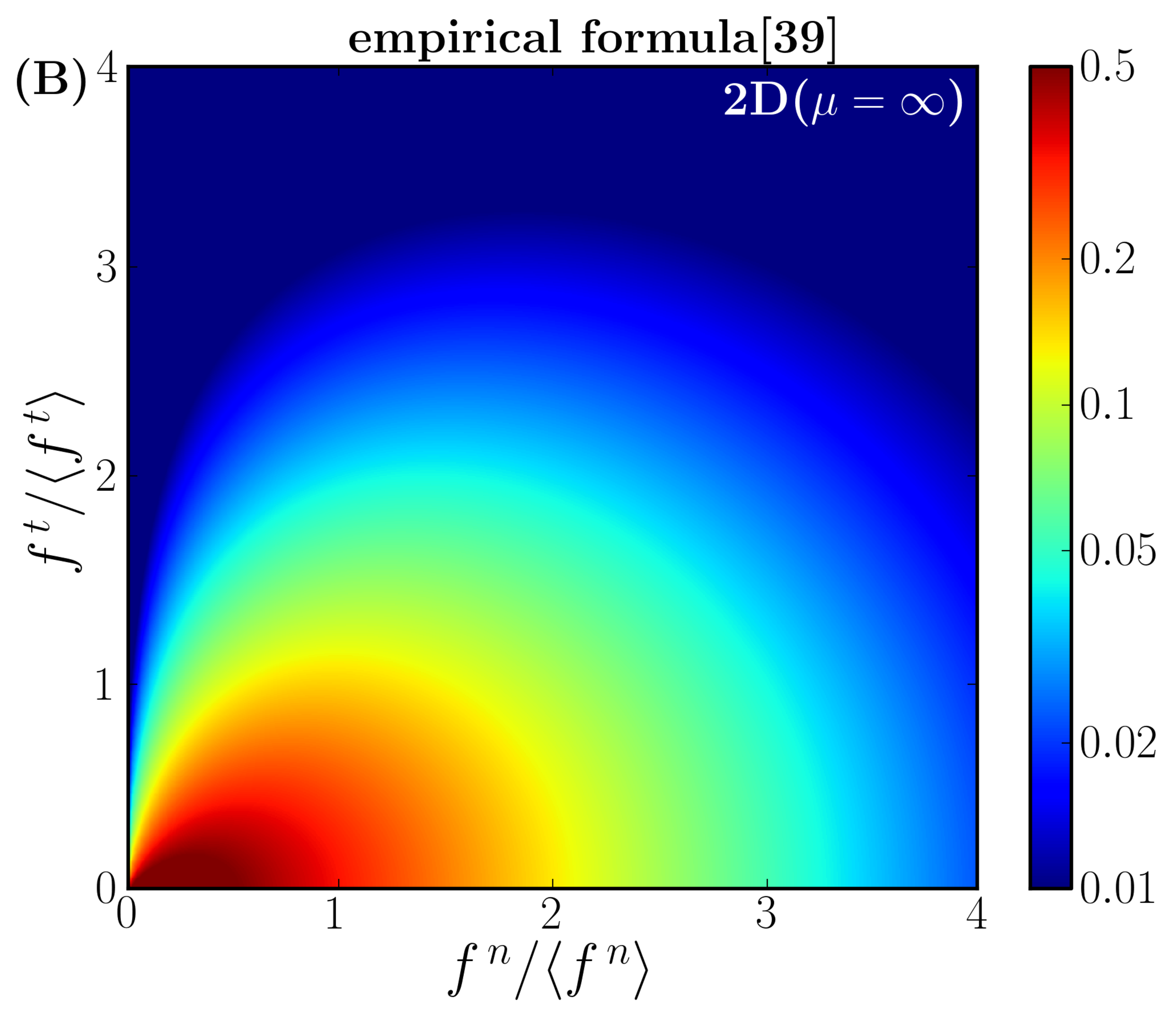}
\includegraphics[width=0.37\textwidth]{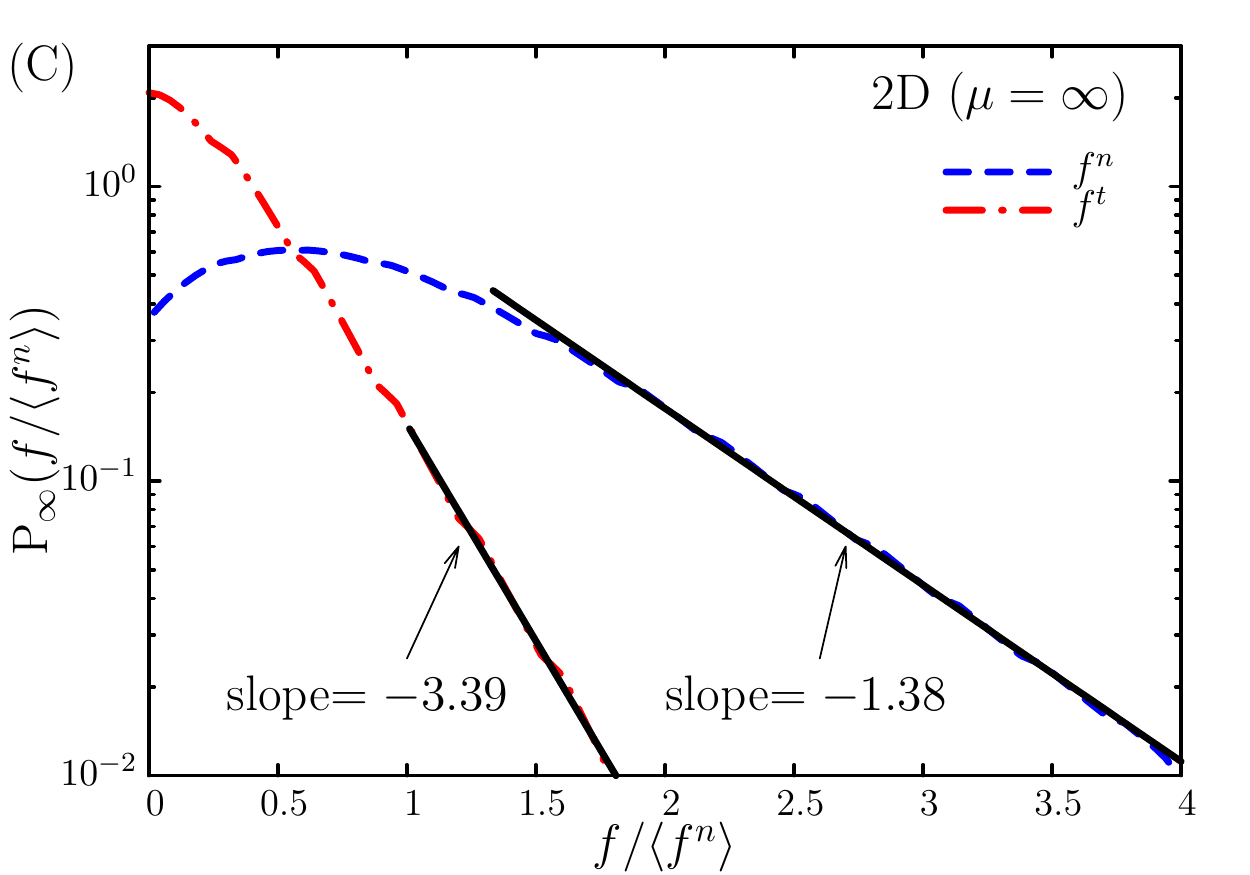}
 \end{minipage} 
\\
\begin{minipage}{1\textwidth}
\includegraphics[width=0.3\textwidth]{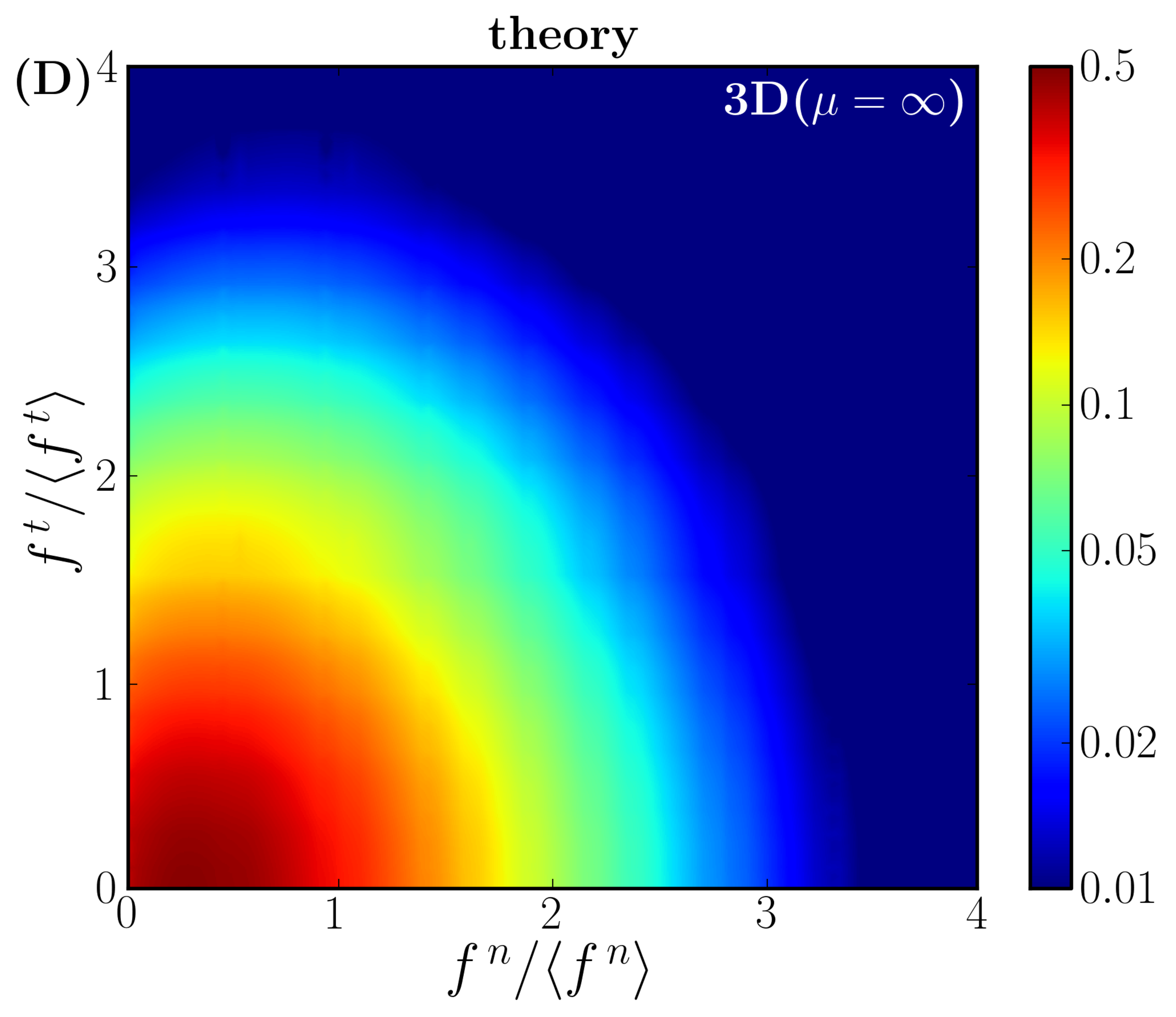}
\includegraphics[width=0.3\textwidth]{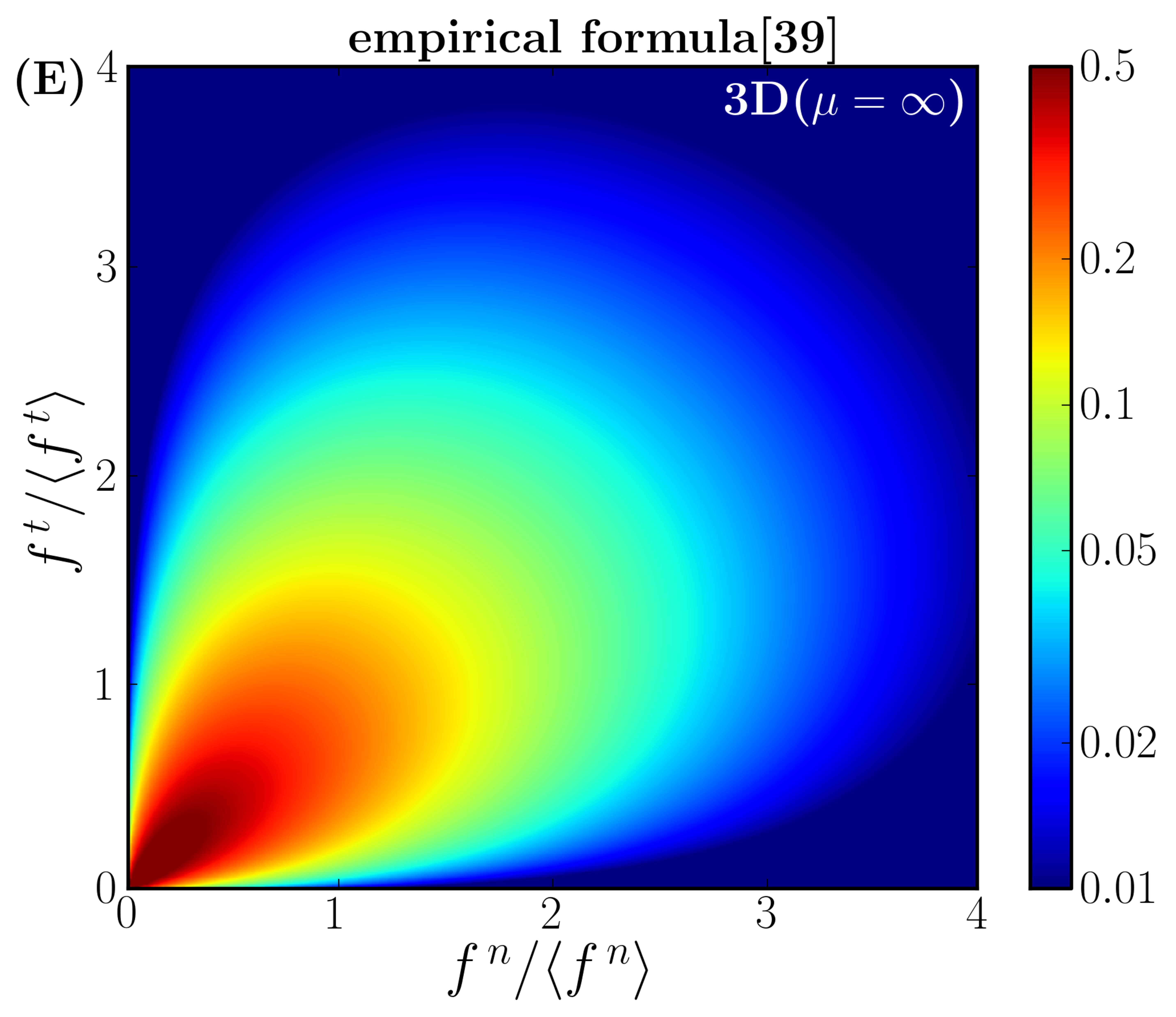}
 \includegraphics[width=0.37\textwidth]{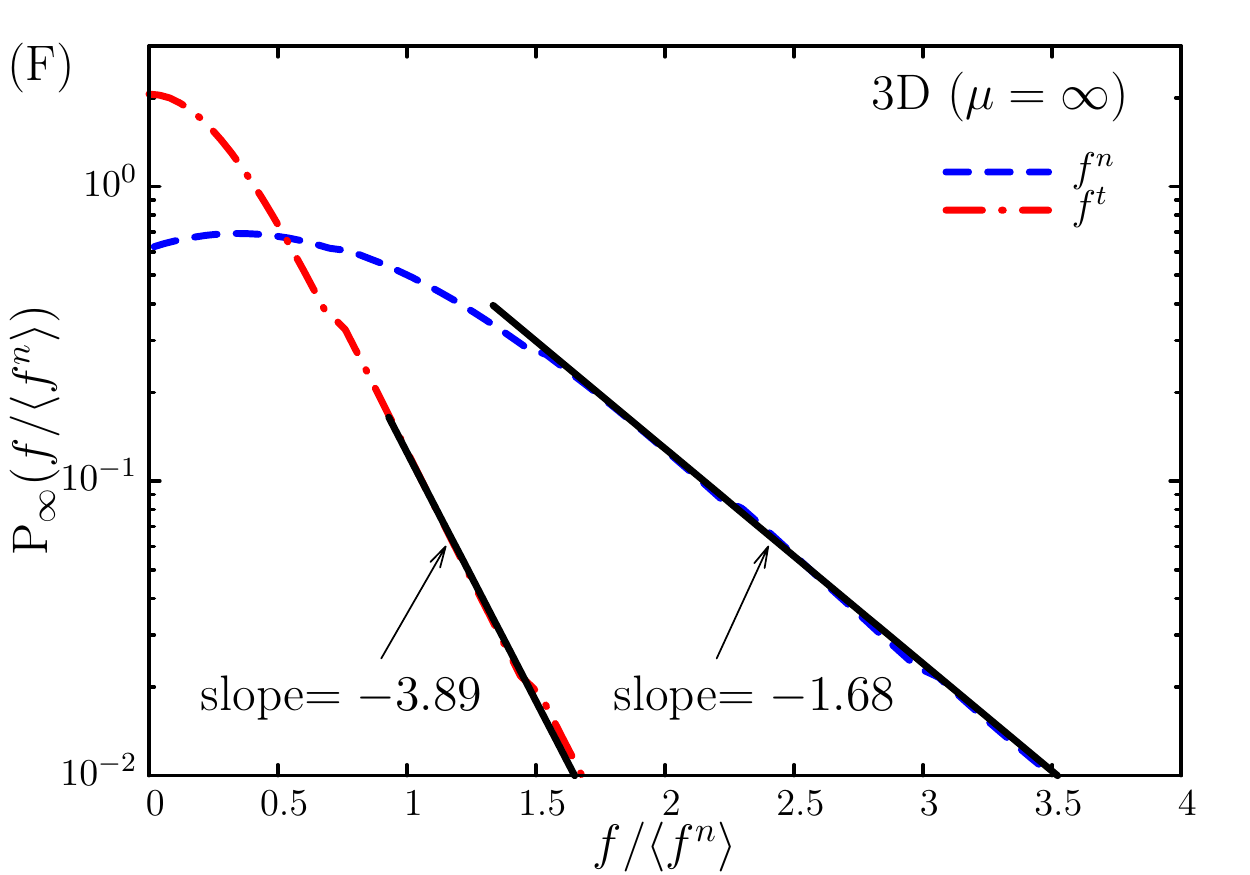}
\end{minipage}
\caption{{\bf Theory and Simulations.} The joint force distribution $\text{P}_{\mu}(f^{\,n}, f^{\,t})$ from theoretical calculation in (A) 2-D spheres packing with $\mu = \infty$, and in (D) 3-D spheres packing with $\mu = \infty$. The contour plots of the empirical formula $\text{P}_{\infty}(f,\theta)=ag(\theta)\exp{(-\sqrt{a}f)}$ with $a=0.8$ which was previously found by Wang {\it et al}~\cite{Wang20103972} are shown in (B) and (E) for 2-D and 3-D respectively. Plots of the corresponding probability distribution of normalized normal forces and frictional forces in 2-D and 3-D are shown in (C) and (F) respectively.
}
\label{fig:pff}
\end{figure*}

\begin{figure}[b]
    \includegraphics[width=0.4\textwidth]{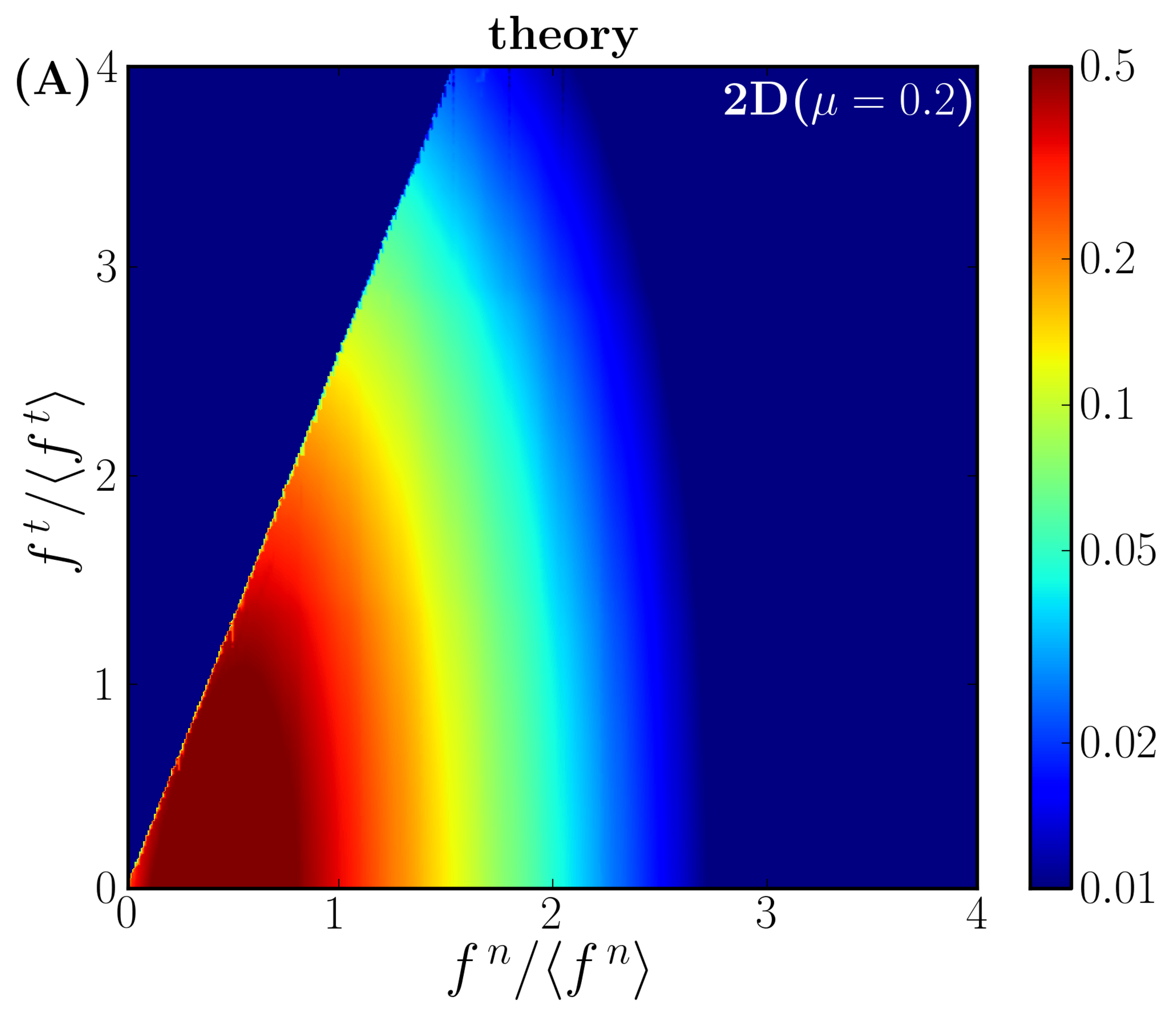}
    \includegraphics[width=0.48\textwidth]{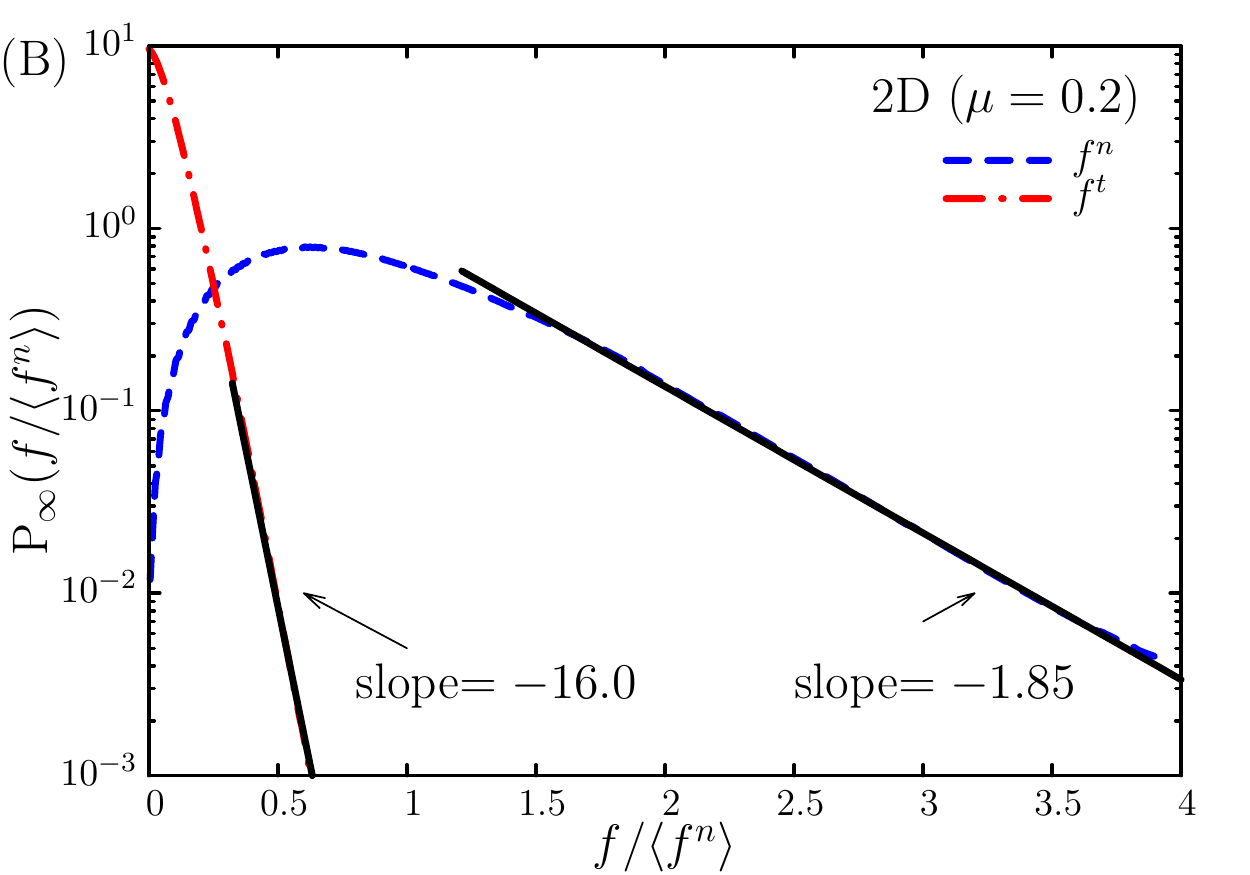}
   \caption{(A) The joint force distribution $\text{P}_{0.2}(f^{\,n}, f^{\,t})$ with friction coefficient $\mu = 0.2$ in 2-D spheres packing. (B) Plots of the probability distribution of normalized normal forces and frictional forces.
   }\label{fig:pff02}
\end{figure}

Similar to the frictionless case, the cavity method can generate the
joint force distribution $\text{P}_{\mu} (f^{\,n}, f^{\,t})$ for frictional 
spheres packings with friction coefficient $\mu$.
The simplest case of infinite friction, $\text{P}_{\mu=\infty}(f^{\,n},
f^{\,t})$ for 2-D and 3-D sphere packings are shown in \reffig{fig:pff}A and \reffig{fig:pff}D respectively, and results follow similar behavior as the empirical fitting formula $\text{P}_{\infty}(f,\theta)=ag(\theta)\exp{(-\sqrt{a}f)}$, (where $g(\theta)=(D-1)(\sin{\theta})^{D-2}\cos{\theta}$, $f\!\!\!=\!\!\!\sqrt{(f^{\,n})^2+(f^{\,t})^2}$, $\theta\!=\!\arctan{(\frac{f^{\,t}}{f^{\,n}})}$ and $a=0.8$) measured in previous numerical studies~\cite{Wang20103972} (\reffig{fig:pff}B and \reffig{fig:pff}E).  In particular, we recover the
non-trivial qualitative change between
2-D and 3-D~\cite{Wang20103972}: while in 3-D, $\text{P}_{\mu=\infty}(f^{\,n}, f^{\,t})
\simeq \text{P}_{\mu=\infty}(f^{\,t}, f^{\,n})$, in 2-D, this symmetry is
clearly broken. This is a consequence of
the more symmetrical role tangential and normal forces play in
3-D with as many torque balance as
force balance equations, whereas in 2-D, there are twice less torque
balance than force balance equations.
When the friction coefficient is finite (see \reffig{fig:pff02}A), the pattern inside the Coulomb cone looks similar to the one obtained at infinite friction. Quite interestingly, we do not
observe any excess of forces at the Coulomb threshold $f^{\,t}= \mu f^{\,n}$,
implying that there are no sliding contacts. This offers a 
theoretical explanation to a singular fact already observed in
simulations: control parameters (essentially volume fraction and friction
coefficient) being equal, the percentage of
plastic contacts in a packing depends on the preparation
protocol~\cite{PhysRevE.72.011301, 0295-5075-101-4-44006}. Our formalism takes into account
those different packings (hence protocols) by performing a
statistical average over possible packings, and the outcome shows that
packings without plastic contacts are dominant. In this regards, the 
fragility associated with the large number of plastic contacts in many
experimentally or numerically generated packings could be mostly
attributed to the preparation protocol, rather than to an inherent
property of random packings of frictional spheres.

Furthermore, we obtain the distributions of the normal
and tangential components for frictionless and frictional packings
as plotted in \reffig{fig:pff}C, \reffig{fig:pff}F and \reffig{fig:pff02}B. 
The normal force distributions all have slight peaks around the mean
 and approximate exponential tails at large forces. Below the mean, 
 the normal force distribution for infinite friction has a nonzero
 probability at zero force whereas it shows a dip towards zero for
 $\mu=0.2$. The tangential force distribution also has an approximate exponential tail, however,
 it decreases monotonically without an obvious rise at small forces. Our results of the probability distribution of normal forces and tangential forces agree with previous experimental measurements in 2-D frictional spheres packing \cite{Majmudar2005}.

\section{Discussion and Conclusion}
In conclusion, we develop a theoretical framework by
using the cavity method, introduced initially for the study of spin-glasses and
optimization problems, to obtain
a statistical physics mean-field description of the force and torque balances
constraints in a random packing. This allows us to get the force
distribution 
and the lower bound on the average coordination number in
frictional and frictionless spheres packings. 

We find a mean-field signature of jamming in the finite value
$\text{P}(f=0)$ of the force distribution at small force.  We also
notice that there is a power law rise $\text P(f)\sim f^{0.52}$ in the
intermediate region of $\text P(f)$.  Thus it is likely that one
obtains an exponent $0 < \theta <0.52$ if the simulations or
experiments can not achieve data down to low enough forces. However,
we must stress that the mean-field approach we develop does not include the detailed
structure of finite dimensional systems, which may have an
important effect on the $\theta$ exponent. Indeed some recent data
seem to extend beyond the intermediate regime $\text P(f)\sim
f^{0.52}$ and find a finite exponent at small
force~\cite{PhysRevLett.109.205501,lerner2013low}.
For frictional packings, we can access the complete joint distribution
$\text P_{\mu}(f^{\,n}, f^{\,t})$.

Concerning the average coordination number, we describe its lower bound
$\zcmm$, which interpolates smoothly between the isostatic frictionless
case $\zcm(0)=2D$, and a large $\mu$ limit $\zc^{\rm
  min}(\infty)$ slightly above $D+1$. This confirms that there is no
discontinuity at $\mu=0$. We predict two scalings for
small and large friction coefficients as $\zcm(0)-\zcmm \sim
{\mu}^{1.03}$ and $\zcmm - \zc^{\rm
  min}(\infty) \sim {\mu}^{-0.36}$.

The statistical mechanics point of view on force and torque balances for random
packings thus proves fruitful. Many features of these systems can be
inferred from those simple considerations. The use of the cavity technique
enables us to tackle this problem with a correct
treatment of the disorder, leading to several new results. This
formalism will be extended to packings of more general shapes and
could be used
to predict other properties of disordered packings, like the yield
stress for instance. Granular materials are not the only systems
subject to force and torque balances, and this constraint is seen in all
overdamped systems, among which suspensions at low Reynolds number constitute
an important example, both conceptually and practically. We hope that
our results will motivate investigations in this direction.

\section*{Appendix: Statistical physics description}
Here we give the statistical physics description of the packing problem
motivating the definition of the partition function $Z$ in \refeq{eq:partition_function} of the
main text.

We define for a given contact network $G$, of $N$ particles and $M$ forces $\vec{f}^{\,a}_i= -
f_{i}^{\,n} \hat{n}^{\,a}_i + f_{i}^{\,t} \hat{t}^{\,a}_i$ exerted on particle $a$ at position $\vec{r}_i^{\,a}$ (with respect to the center of the particle), an energy function which is the sum of the square of total forces and torques on each particle,
\begin{equation}
\label{H}
H = \sum^N_{a=1}\Big( \sum_{i \in \partial a}\vec f_{i}^{\,a} \Big)^2 + \sum^N_{a=1}\Big( \sum_{i \in \partial a} \vec r_i^{\,a} \times \vec f_{i}^{\,a} \Big)^2 + \Big(\sum_{i=1}^M f_{i}^{\,n} - Mp\Big)^2.
\end{equation}

We then define the partition function of contact network $G$ with quenched disorder on the contact normals $\{\hat{n}^{\,a}_i\}$,
\begin{equation}
\begin{array}{rl}
 Z_{\beta} = & \!\!\! \displaystyle \int \prod_{i=1}^M \Big[ \dd f_{i}^{\,n} \dd f_{i}^{\,t} 
   \dd  \hat{t}^{\,a}_i \dd \hat{t}^{\,b}_i  \, \delta(\hat{t}^{\,a}_i
   + \hat{t}^{\,b}_i)\\
   & \qquad\qquad\qquad \displaystyle \Theta(f_{i}^{\,n}) \Theta(\mu
  f_{i}^{\,n} - f_{i}^{\,t}) \Big] \mathrm{e}^{-\beta H},
  \end{array}
\end{equation}
where $\beta$ is an inverse temperature, which is only here as a
parameter (it is not the actual temperature, which is irrelevant for
granular packings), and will be set to $\beta \to \infty$ soon. The
Heaviside $\Theta$ functions ensure repulsive normal forces, and
Coulomb friction condition, respectively.

The ground state of the hamiltonian $H$ at $\beta \to \infty$ provides the balance condition in the packing. Consider the repulsive nature of contact forces and the Coulomb condition when friction exists, by using the factor graph representation of the Boltzmann probability, the weight of interaction node $a$ (a particle on which force/torque balance is satisfied) is 
\begin{equation}
  \begin{array}{rl} 
 & \hskip-0.2in \chi_a^{\beta}(\{ f^{\,n},f^{\,t},\hat{n}^{\,a},\hat{t}^{\,a} \}_{\partial
  a}) \displaystyle = \frac{1}{Z^{\,a}_{\beta}} \exp\Big[-\beta \big(\displaystyle\sum_{i\in\partial a}\vec f_i^{\,a} \big)^2 \\
& -\beta \big(\displaystyle\sum_{i\in\partial a}\vec r_i^{\,a} \times \vec f_{i}^{\,a} \big)^2
 \Big] \displaystyle \prod_{i \in \partial a} \Theta(f_{i}^{\,n}) \Theta(\mu
  f_{i}^{\,n} - f_{i}^{\,n}) ,
  \end{array}
\end{equation}
with $Z^{\,a}_{\beta}$ ensuring \[1=\int \prod_{i \in \partial a}
 \dd f_{i}^{\,n} \dd f_{i}^{\,t} 
   \dd  \hat{t}^{\,a}_i   \chi_a^{\beta}(\{
f^{\,n},f^{\,t},\hat{n}^{\,a},\hat{t}^{\,a} \}_{\partial a}) .\]
  In the zero temperature limit $\beta \to \infty$, the above expression becomes the
force/torque balance constraint on each node:
\begin{equation}
\begin{array}{rl}
& \hskip-0.2in \chi_a(f) = \displaystyle \lim_{\beta \to \infty} \chi_a^{\beta}(f) \\
& \hskip-0.1in = \delta \Big(\displaystyle\sum_{i\in\partial a}\vec f_i^{\,a} \Big) 
\delta \Big(\displaystyle \sum_{i\in\partial a}\vec r_i^{\,a} \times \vec f_{i}^{\,a} \Big)
 \displaystyle \prod_{i \in \partial a} \Theta(f_{i}^{\,n}) \Theta(\mu
  f_{i}^{\,n} - f_{i}^{\,t}),
  \end{array}
\end{equation}
Notice that in this limit the exact shape of hamiltonian Eq.~(\ref{H})
is irrelevant, as its only condition is that it provides the
force and torque balances in the limiting case $\beta \to \infty$.

The associated partition function $Z = \lim_{\beta \to \infty}
Z_{\beta}$ is then the one defined in \refeq{eq:partition_function} of the main text.
This partition function is defined at the level of a single graph, but
we study the problem for an ensemble of random graphs, \textit{ie} we
average the entropy $\log Z$ over the selected ensemble of random
graphs, defined by the
distribution of connectivity (contact number) $\mathrm{R}(z)$ and the
joint probability distribution of the contact directions
$\Omega(\{\hat{n}\})$ around every interaction node (particle),
as explained in the main text. This ensemble is . 

\section*{Acknowledgments}
We gratefully acknowledge funding by NSF-CMMT and DOE Office of Basic
Energy Sciences, Chemical Sciences, 
Geosciences, and Biosciences Division. We thank
F. Krzakala and Y. Jin for interesting discussions.

\footnotesize{

\providecommand*{\mcitethebibliography}{\thebibliography}
\csname @ifundefined\endcsname{endmcitethebibliography}
{\let\endmcitethebibliography\endthebibliography}{}

}

\end{document}